\documentclass[reprint, floatfix, superscriptaddress, aip, jcp]{revtex4-1}
\usepackage{amsmath,amssymb}
\usepackage{graphicx,hyperref,color}

\newcommand{\nn}{\nonumber}
\newcommand{\dunder}[1]{\underline{\underline{#1}}}
\begin{document}
\title{Density functional approach to elastic properties of three-dimensional dipole-spring models for magnetic gels}
\author{Segun Goh}
\email{s.goh@fz-juelich.de}
\affiliation{Theoretical Physics of Living Matter, Institute of Biological Information Processing, Forschungszentrum J{\"u}lich, 52425 J{\"u}lich, Germany}
\affiliation{Institut f\"ur Theoretische Physik II: Weiche Materie, Heinrich-Heine-Universit\"at D\"usseldorf, Universit{\"a}tsstr. 1, 40225 D\"usseldorf, Germany}

\author{Andreas M. Menzel}
\email{a.menzel@ovgu.de}
\affiliation{Institut f\"ur Physik, Otto-von-Guericke-Universit\"at Magdeburg, Universit{\"a}tsplatz 2, 39106 Magdeburg, Germany}

\author{Ren{\'e} Wittmann}
\email{rene.wittmann@hhu.de}
\affiliation{Institut f\"ur Theoretische Physik II: Weiche Materie, Heinrich-Heine-Universit\"at D\"usseldorf, Universit{\"a}tsstr. 1, 40225 D\"usseldorf, Germany}

\author{Hartmut L{\"o}wen}
\email{hlowen@hhu.de}
\affiliation{Institut f\"ur Theoretische Physik II: Weiche Materie, Heinrich-Heine-Universit\"at D\"usseldorf, Universit{\"a}tsstr. 1, 40225 D\"usseldorf, Germany}
\date{\today}
\begin{abstract}
Magnetic gels are composite materials, consisting of a polymer matrix and embedded magnetic particles.
Those are mechanically coupled to each other, giving rise to the magnetostrictive effects
as well as to a controllable overall elasticity responsive to external magnetic fields. 
Due to their inherent composite and thereby multiscale nature, 
a theoretical framework bridging different levels of description is indispensable 
for understanding the magnetomechanical properties of magnetic gels.
In this study, we extend a recently developed density functional approach from two spatial dimensions to more realistic three-dimensional systems.
Along these lines, we connect a mesoscopic characterization resolving the discrete structure of the magnetic particles, to macroscopic continuum parameters of magnetic gels.
In particular, we incorporate 
the long-range nature of the magnetic dipole-dipole interaction,
and consider the approximate incompressibility of the embedding media,
and relative rotations with respect to an external magnetic field breaking rotational symmetry.
We then probe the shape of the model system in its reference state,
confirming the dependence of magnetostrictive effects on the configuration of the magnetic particles and on the shape of the considered sample.
Moreover, calculating the elastic and rotational coefficients on the basis of our mesoscopic approach,
we examine how the macroscopic types of behavior are related to the mesoscopic properties.
Implications for real systems of random particle configurations are also discussed.
\end{abstract}
\maketitle
%
\section{Introduction}
\label{sec:intro}
Ferrogels, magnetic gels as well as magnetorheological gels and elastomers, all referred to as magnetic gels, are 
soft elastic composite materials containing magnetic or magnetizable particles, 
both simply referred to as magnetic particles. 
Their mechanical properties are controllable 
by external magnetic fields~\cite{Filipcsei2007, Ilg2013, Menzel2015, Odenbach2016}.
The composite nature arises as the magnetic particles are mechanically coupled to
a surrounding polymeric matrix~\cite{Frickel2011, Messing2011, Gundermann2014, Landers2015, Roeder2015}.
Such a magnetomechanical coupling has even been enhanced by anchoring polymers directly
on the surface of magnetic particles~\cite{Ilg2013,Frickel2011,Messing2011,Roeder2015}.
To understand the rheological properties of these materials, 
the dependence of their elastic moduli and magnetostrictive effects on external magnetic fields have been investigated 
in various settings~\cite{Ginder2002, Zhou2004, Gollwitzer2008, Borin2013, Kramarenko2015, Safronov2019, Borin2019, Saveliev2020}.
In particular, induced changes in the configuration of the magnetic particles, 
especially the touching of adjacent particles and chain formation,
have been repeatedly reported as prominent features in the response of magnetic gels to external magnetic fields~\cite{Danas2012, Gundermann2014, Gundermann2017, Pessot2018, Puljiz2018, Sturm2019, Schumann2020}. 

Due to their inherent composite nature, 
a complete theoretical understanding of magnetic gels is still
challenging~\cite{Cremer2017}.
At macroscopic scales, thermodynamic and hydrodynamic theories have been developed~\cite{Jarkova2003, Bohlius2004, Potisk2019},
in which both the magnetic and elastic components are modeled as homogeneous continua.
Notably, the positive magnetostriction, namely elongation along the magnetic field direction,
has been predicted using such continuum approaches~\cite{Raikher2005, Raikher2008, Stolbov2019}.
However, potential effects stemming from the detailed configurations of the magnetic particles and the polymers
as well as the mechanisms governing these effects are hardly resolved at this scale.
Rather than that, phenomenological coefficients have to be determined via modeling or experiments.
In a theoretical perspective, one may consider a model 
resolving all the magnetic particles and polymer molecules at a microscopic level.
Indeed, numerical simulation studies have been performed at this scale,
revealing the roles of the polymer network topology and 
the coupling between the orientation of magnetic particles and the surrounding polymers~\cite{Weeber2012, Weeber2015jcp},
as well as the degree of cross-linking in the polymer matrix~\cite{Minina2018}.
However, unifying all the ingredients of such models to derive macroscopic parameters  
remains a demanding task.

In this regard, mesoscopic approaches still address the configurations of the magnetic particles explicitly, while individual polymeric building blocks are not resolved. 
Indeed, the significance of detailed structures at mesoscopic length scales has been revealed,
as the mesoscopic models predict, for instance, both positive and negative magnetostrictive effects
depending on the specific configuration of the magnetic particles~\cite{Stolbov2011, Ivaneyko2012, Fischer2019, Romeis2019}.
The rotational fluctuations of magnetic particles have also been addressed within a mesoscopic approach~\cite{Wood2011}.
Specifically, the polymer matrix can be coarse-grained 
as an elastic continuum~\cite{Biller2014, Biller2015, Puljiz2016, Puljiz2017}.
We note that the role of the magnetic particles can also be modeled 
using continuum fields~\cite{Metsch2016, Romeis2017} 
that describe the particle arrangements. 

Alternatively, the elastic continuum can be discretized on the mesoscopic scale 
as a network of harmonic springs~\cite{Pessot2014, Pessot2016, Menzel2019}.
One advantage of this approach is that microscopic theories as well as simulation techniques 
developed in the framework of statistical mechanics are directly applicable.
The interaction energies are explicitly defined in this case.
As demonstrated in Ref.~\onlinecite{Menzel2014} using a description of a uniaxial magnetic gel, 
and in Refs.~\onlinecite{Cremer2017, Goh2019} using an approach for isotropic one- and two-dimensional systems,
a bridging description between mesoscopic and macroscopic scales
may unravel the role of the discrete mesoscopic structures in the materials for the macroscopic behavior.
In this way, the gap between continuum theories and mesoscopic models is closed.

In the present study, we further explore the issue of scale-bridging and, in particular,
the statistical mechanics of magnetic gels.
As for the mesoscopic description, we employ a simple but tangible model 
consisting of magnetic dipolar particles and harmonic springs connecting them.
Starting from the mesoscale model, 
we aim at calculating macroscopic elastic and rotational coefficients,
the trend of which we then correlate with mesoscopic characteristics, 
i.e., the configuration of the magnetic particles.
Specifically, 
we employ classical density functional theory (DFT)~\cite{Evans1979, Lowen2002, Oxtoby2002, Evans2016},
extending the concept of mapping the elastic interactions 
between the particles through the surrounding elastic medium 
onto pseudosprings~\cite{Cremer2017, Goh2019} to three dimensions.
The resulting free energy allows for a calculation of macroscopic elasticity parameters. 

To this end, the following issues need to be addressed in advance.
First of all, the dipolar magnetic interaction is strictly long-ranged, 
rendering the system thermodynamically ill-defined~\cite{Dauxois2002}.
While the Ewald summation technique~\cite{Arnold2005, Allen2017} 
can be adopted to numerically simulate systems with long-range interactions such as suspensions of magnetic particles in liquid crystalline matrices~\cite{Peroukidis2016,Siboni2020},
the shape dependence of the free-energy has to be clarified 
as in the studies of dipolar fluids~\cite{Groh1994prl, Groh1994pre} and of magneto-sensitive elastomers~\cite{Ivaneyko2014}.
In addition, we note that the aforementioned magnetostrictive effects originate from 
the anisotropic nature of the magnetic dipole-dipole interaction.
These points require a careful choice of the DFT implementation. 
Second, thermal fluctuations of the magnetic particles need to be included.
As we develop a statistical theory for the equilibrium free energy, i.e., DFT, this issue is resolved automatically.
Third, when an external magnetic field explicitly breaks the rotational symmetry of the system,
relative rotations with respect to the field direction should be considered in the underlying elasticity theory.
Originally, this concept was introduced in the context of liquid-crystalline elastomers~\cite{deGennes1980,Brand1994,Menzel2007,Menzel2009},
but has also been extended to uniaxial magnetic gels~\cite{Bohlius2004, Menzel2014}.
Lastly, the role of incompressibility that may be inherent in many systems of magnetic gels should also be clarified in the description.
Just as conventional gels, magnetic gels can swell/shrink by absorbing/releasing liquid.
Otherwise, they are regarded as incompressible, for instance, due to the dispersed fluid.
Such incompressibility may alter the mechanical properties of magnetic gels~\cite{Filipcsei2010,Safronov2019},
calling for a theory respecting volume conservation.

This paper is organized as follows.
In Sec.~\ref{sec:model} our model for magnetic gels is introduced.
We then formulate elasticity theory for incompressible systems 
in Sec.~\ref{sec:elasticity}, including the components of relative rotations.
The DFT for our model system and its implementation are described in Sec.~\ref{sec:DFT}.
In Sec.~\ref{sec:result}, 
we present the elastic and rotation coefficients as well as magnetostrictive effects 
obtained from the DFT calculation.
Lastly, discussions are included in Sec.~\ref{sec:discussion}.
%
\section{Mesoscopic dipole-spring model}
\label{sec:model}
We consider a three-dimensional version of 
the previously studied dipole-spring system~\cite{Annunziata2013, Pessot2016}
as a mesoscopic model for magnetic gels.
The model consists of $N$ identical magnetic particles 
of diameter $\sigma_D$ and dipole moment $\mathbf{m}$,
which are connected by identical harmonic springs of spring constant $k_{\rm el}$ and rest length $a_{\rm el}$.
The position of the $i$th particle is denoted by $\mathbf{r}_i$ ($i=1,\ldots,N$). 

The total Hamiltonian $\mathcal{H}_{\rm tot}$ of the system is introduced as the sum of
the kinetic part $\mathcal{H}_{\rm kin}$ and the interaction part $\mathcal{H}_{\rm int}$ 
of the magnetic particles, 
the latter of which splits into three parts:
\begin{align}
\mathcal{H}_{\rm int} = \mathcal{H}_{\rm m} + \mathcal{H}_{\rm el} + \mathcal{H}_{\rm st}.
\label{eq_Hint}
\end{align}
Here, $\mathcal{H}_{\rm m}$ and $\mathcal{H}_{\rm st}$,
respectively, denote the energies of the magnetic dipole-dipole interaction and the steric repulsion, 
which are all-to-all pairwise additive. 
Therefore, with $\mathbf{r}_{ij} = \mathbf{r}_j - \mathbf{r}_i$, they take the form 
\begin{align}\label{eq:pairadd}
\mathcal{H}_{\rm m, st} = \frac{1}{2} \sum_{i\neq j} u_{\rm m, st} (\mathbf{r}_{ij}),
\end{align}
where $u_{\rm m}$ and $u_{\rm st}$ denote the two-body magnetic dipole-dipole interaction and steric repulsion,
respectively, as detailed below.

In stark contrast to Eq.~\eqref{eq:pairadd}, the elastic part $\mathcal{H}_{\rm el}$ does not simply take the form of a pairwise additive potential,
namely no general two-body potential applying simultaneously to all pairs of particles can be introduced.
Specifically, the elastic contribution is written in the form
\begin{align}
\mathcal{H}_{\rm el} = \sum_{\langle i, j \rangle} u_{\rm el}(r_{ij}) =\sum_{\langle i, j \rangle} \frac{1}{2} k_{\rm el}({r}_{ij} -a_{\rm el})^2,
\label{eq_Hel}
\end{align}
where $\langle i, j \rangle$ indicates that the sum only includes a predefined set of neighbors, which labels the particles such that they become distinguishable. 
Thus the potential energy cannot be written as a sum over pair potentials of indistinguishable particles. 
For the two-body potential $u_{\rm el}$ a harmonic spring potential of spring constant $k_{\rm el}$ is adopted,
while $a_{\rm el}$ is the rest length of the springs and $r_{ij} =|\mathbf{r}_{ij}|$. 
Here, we assume a face-centered cubic (FCC) lattice structure with twelve nearest-neighbor particles,
which is indicated by the angular bracket in Eq.~\eqref{eq_Hel}.
Therefore, in total $6N$ harmonic springs connect the nearest-neighboring pairs of magnetic particles in this specific model (except for boundary effects). 

Next, for the two-body steric repulsion in Eq.~\eqref{eq:pairadd}, we assume a particle diameter $\sigma_D$ and adopt a hard-core potential in the form of
\begin{equation}
\label{eq_Hst}
u_{\rm st}(\mathbf{r}) =\begin{cases}
	0 & \ {\rm if}  \ r \geq \sigma_D , \\
	\infty & \ {\rm otherwise.}
\end{cases}
\end{equation}
As a dimensionless density we introduce 
the packing fraction $\eta$
defined as the fraction of the volume occupied by the magnetic particles,
i.e., $\eta \equiv N(4\pi/3)(\sigma_D/2)^3/V$ where $V$ is the volume of the system. 

\begin{figure}
\includegraphics[width=8.5cm]{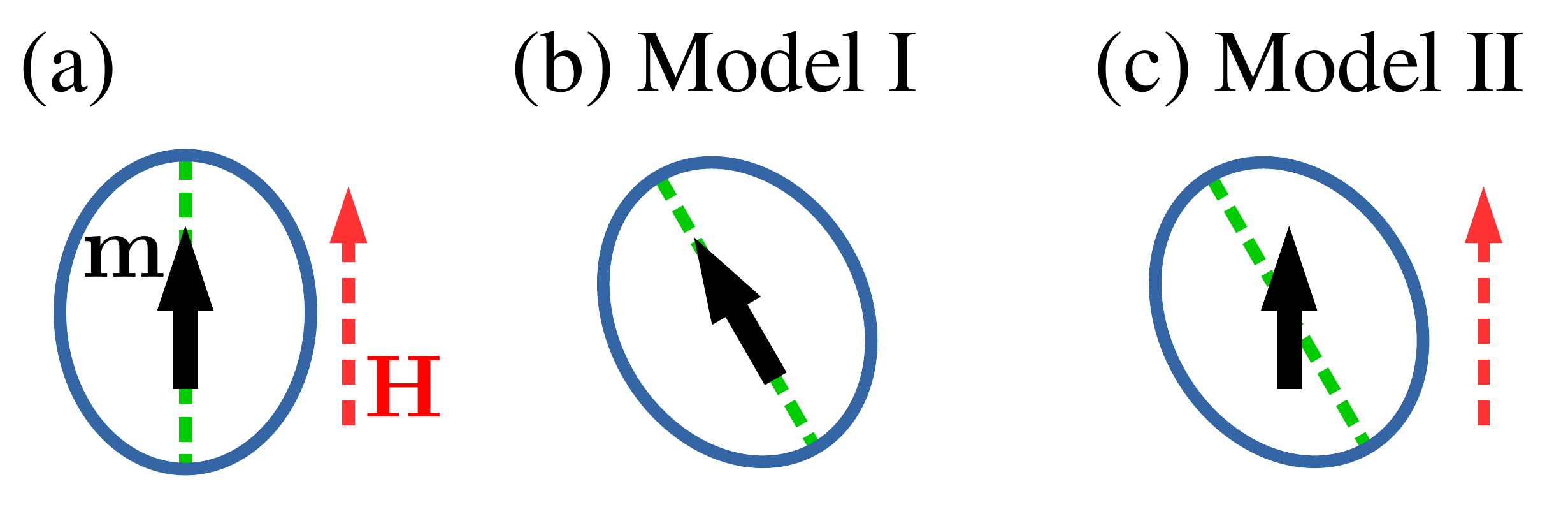} 
\caption{\label{fig:model}
(a) Under an external magnetic field $\mathbf{H}$, 
the magnetic dipole moment $\mathbf{m}$ (black solid arrow)
is aligned along the external field direction $\mathbf{H}$ (red dashed arrow).
Here the whole system (solid ellipse) has the same orientation
(green dashed line at the center of the ellipse) as the initial magnetization, 
which is the reference state of our model system.
Regarding rotations, we consider two models, (b) Model I for ferromagnetic particles and 
(c) Model II for paramagnetic particles, see text for details.} 
\end{figure}

Specifying the two-body magnetic dipole-dipole interaction energy, the two-body potential in Eq.~\eqref{eq:pairadd} reads
\begin{align}
u_{\rm m}(\mathbf{r}) &= \frac{\mu_0}{4\pi} 
\left[ \frac{\mathbf{m}\cdot \mathbf{m}}{r^3}-
\frac{3(\mathbf{m}\cdot \mathbf{r})(\mathbf{m}\cdot \mathbf{r})}{r^5} \right],
\label{eq:u_m} 
\end{align}
where $\mu_0$ is the vacuum permeability.
The magnetic moment $\mathbf{m}$ is determined by the applied magnetic field $\mathbf{H}=H\hat{z}$, which is always directed along the $z$-direction, and
the magnetization properties of each magnetic particle, see, e.g., Refs.~\onlinecite{Sanchez2019, Vaganov2020}.
When an external magnetic field is applied, we assume that $\mathbf{m}(\mathbf{H}) \parallel \mathbf{H}$,
i.e., $\mathbf{m} = m\hat{z}$, see Fig.~\ref{fig:model}(a) for illustration.
In the absence of the applied field, magnetic particles may or may not retain their magnetization. 
Here we consider two simple cases.
First, in Model I, we assume that the magnetic particles are ferromagnetic.
There, the magnetic moment of each particle persists and is fixed with respect to the particle orientation, once the magnetic particles are magnetized.
We then investigate elastic properties of the model system in the absence of external fields as depicted in Fig.~\ref{fig:model}(b).
In this case, the dipole moment $\mathbf{m}$ rotates rigidly with the whole system.
In Model II, we assume that the magnetic particles are paramagnetic. 
To retain the magnetization, the external field $\mathbf{H}$ needs to be persistently applied to the system in this case.
In contrast to Model I, $\mathbf{m}$ is then always directed along $\mathbf{H}$ as shown in Fig.~\ref{fig:model}(c),
even when the whole system rotates.
Consequently, relative rotations between the magnetization direction 
and the rest of the system become relevant.
We note that the magnetic dipole-dipole interaction breaks the isotropy of the system both in Model I and Model II.

As for the orientation of the system, we consider two cases in which 
the $(0,0,1)$- and $(1,1,1)$-orientations of the FCC lattice
are directed along the $z$-axis~\cite{Fischer2019}.
When the lattices are elongated or contracted along the $z$-direction 
due to the anisotropic magnetic interaction, 
the resultant lattices of the $(0,0,1)$- and $(1,1,1)$-cases
are tetragonal and rhombohedral, respectively.

Lastly, we assume that our model system is incompressible, i.e.,
the volume of the whole system is fixed and persists even under deformations. 
Here we set $V = (\sqrt{2}/2)a_{\rm el}^3 N$, at which the total Hamiltonian 
$\mathcal{H}_{\rm tot}$ is minimized for $m = 0$. 

\section{Macroscopic description}
\label{sec:elasticity}

As incompressibility is assumed in our model system,
we should address the maintained volume conservation
when developing our macroscopic description.
While our model system is initialized with the prescribed volume $V$ at $m=0$,
an external magnetic field induces a magnetostrictive effect,
which is not necessarily volume preserving.
However, the imposed incompressibility constraint 
hinders the system to relax to a new volume upon magnetization.  
Such effects introduce a predeformation hidden behind the maintained volume, 
rendering our model system nonlinear elastic.
Here, following the group theoretical approach proposed in Ref.~\onlinecite{Goh2022},
we consider nonlinear elastic responses of our model system
and calculate elastic moduli accordingly.
For self-containedness, we briefly summarize the formulation and 
introduce the second-order corrections to the deformation gradient tensors 
that are relevant for the computation of elastic moduli.

\subsection{Nonlinear deformation gradient tensor}
\label{sec:group}
Under the incompressibility condition, the deformation gradients in three dimensions are elements of the special linear group $\mathsf{SL}(3,\mathbb{R})$, the Lie algebra of which
is $\mathfrak{sl}(3,\mathbb{R})$.
Generally, the components of the deformation gradient tensor $\mathbf{F}$
are defined by 
$\mathbf{F}_{ij}=\partial \mathbf{r}'_i/\partial \mathbf{r}_j$, 
where $\mathbf{r}'$ and $\mathbf{r}$ mark the positions of the material elements in the deformed and undeformed state, respectively. 
Then nonlinear deformation gradient tensors $\mathbf{F}$ may be expressed
via the exponential map 
\begin{align} \label{eq:DGT}
\mathbf{F} = 
\exp{\left( \sum_{i=1}^8 \epsilon_i \boldsymbol{\lambda}_i \right)},
\end{align}
where, $\boldsymbol{\lambda}_i$ denote the $\mathsf{SL}(3,\mathbb{R})$ group generators 
and $\epsilon_i$ are small coefficients indicating the magnitude of deformations generated by $\boldsymbol{\lambda}_i$. 
One should choose a set of generators, which is appropriate for the system considered. 
Accordingly, for our model system, we employ the generators of
\begin{align} \label{eq:SL3_generators}
&\boldsymbol{\lambda}_1 = \left( \begin{array}{ccc}
1 & 0 & 0 \\ 0 & -1 & 0 \\ 0 & 0 & 0  \end{array} \right), \quad
\boldsymbol{\lambda}_2 = \frac{1}{\sqrt{3}}\left( \begin{array}{ccc}
1 & 0 & 0 \\ 0 & 1 & 0 \\ 0 & 0 & -2  \end{array} \right), \nonumber \\ 
&\boldsymbol{\lambda}_3 = \left( \begin{array}{ccc}
0 & 0 & 0 \\ 0 & 0 & 1 \\ 0 & 1 & 0  \end{array} \right), \quad
\boldsymbol{\lambda}_4 = \left( \begin{array}{ccc}
0 & 0 & 1 \\ 0 & 0 & 0 \\ 1 & 0 & 0  \end{array} \right), \nonumber \\
&\boldsymbol{\lambda}_5 = \left( \begin{array}{ccc}
0 & 1 & 0 \\ 1 & 0 & 0 \\ 0 & 0 & 0  \end{array} \right), \quad
\boldsymbol{\lambda}_6 = \left( \begin{array}{ccc}
0 & 0 & 0 \\ 0 & 0 & -1 \\ 0 & 1 & 0  \end{array} \right),\nonumber \\
&\boldsymbol{\lambda}_7 = \left( \begin{array}{ccc}
0 & 0 & 1 \\ 0 & 0 & 0 \\ -1 & 0 & 0  \end{array} \right), \quad
\boldsymbol{\lambda}_8 = \left( \begin{array}{ccc}
0 & -1 & 0 \\ 1 & 0 & 0 \\ 0 & 0 & 0  \end{array} \right). 
\end{align}
Here, the transformations associated with $\boldsymbol{\lambda}_1$ stretch (compress) the system along the $x$-axis,
combined with compressions (stretches) along the $y$-axis;
the deformations generated by $\boldsymbol{\lambda}_2$ stretch (compress) the system in the $xy$-plane combined with compressions (stretches) along the $z$-axis of twice the magnitude;
$\boldsymbol{\lambda}_3$, $\boldsymbol{\lambda}_4$, and $\boldsymbol{\lambda}_5$ generate shear deformations in the $yz$-, $zy$-, and $xy$-plane, respectively;
$\boldsymbol{\lambda}_6$, $\boldsymbol{\lambda}_7$, and $\boldsymbol{\lambda}_8$ generate rotations in the $yz$-, $zy$-, and $xy$-plane, respectively. 
We note that $\boldsymbol{\lambda}_5$ can also be regarded as a generator of shear deformations
in the $xy$-plane, but with orientations different from those generated by $\boldsymbol{\lambda}_1$.
For the purpose of calculating elastic moduli,
corrections up to the second order of $\epsilon_i$ are relevant.
Accordingly, we may truncate the expansion at the third order of $\{\epsilon_i\}$ and use
\begin{align} \label{eq:nonlinear_F}
\mathbf{F} = \mathbf{I} + \sum_i \epsilon_i \boldsymbol{\lambda}_i 
+\frac{1}{2}\left( \sum_i \epsilon_i \boldsymbol{\lambda}_i \right)\cdot\left( \sum_i \epsilon_i \boldsymbol{\lambda}_i \right).
\end{align}
We note that, in general, generators do not commute, i.e., 
$\boldsymbol{\lambda}_i \cdot \boldsymbol{\lambda}_j \neq \boldsymbol{\lambda}_j \cdot \boldsymbol{\lambda}_i$. 
Within our approach, the free-energy density $\mathcal{F}$ 
(see Sec.~\ref{sec:DFT} for the definition based on density functional theory),
equivalent to the deformation energy density in nonlinear elasticity,
is regarded as a function of $\{\epsilon_i\}$.
This choice naturally allows us to define the generalized elastic moduli as 
\begin{align} \label{eq:generalized_C_definition}
C_{ij} = \frac{\partial^2 \mathcal{F}}{\partial \epsilon_i \partial \epsilon_j}.
\end{align}

In the case of Model I, only the five generators $\boldsymbol{\lambda}_i$ for $i=1,\ldots,5$ are relevant,
among which the shear deformations generated by $\boldsymbol{\lambda}_3$ and $\boldsymbol{\lambda}_4$ 
lead to identical contributions to $\mathbf{F}$ due to the symmetry of tetragonal and rhombohedral lattices.
In addition to those, the relative rotations corresponding to $\boldsymbol{\lambda}_6$
and $\boldsymbol{\lambda}_7$ must be included for the description of Model II,
whereas rotations in the $xy$-plane generated by $\boldsymbol{\lambda}_8$ are still irrelevant. 
Again due to the symmetry, the rotations corresponding to $\boldsymbol{\lambda}_6$
and $\boldsymbol{\lambda}_7$ lead to identical contributions to $\mathbf{F}$. 

\subsection{Irreducible representation for stiffness tensors}
\label{sec:irreducible}
In the framework of linear elasticity theory, 
irreducible representations for stiffness tensors are determined 
by the underlying symmetry of the systems.
We here consider the strains which are defined in linear elasticity as
\begin{align} \label{eq:linear_strain}
\epsilon_{ij} \equiv \frac{1}{2}(\mathbf{\nabla}_i \mathbf{u}_j +\mathbf{\nabla}_j \mathbf{u}_i),
\end{align}
where $\mathbf{u}$ denotes the displacement field.
For a tetragonal lattice [$(0,0,1)$-orientation], 
the stiffness matrix $\dunder{C}$ in Mandel (or orthonormal) notation, 
where the stiffness matrix becomes a second-rank tensor~\cite{Mehrabadi1990,Mazdziarz2019}, takes the form
\begin{align}
\dunder{C}_{\rm Mandel}^{\rm Tetr} = \left( \begin{array}{cccccc}
\tilde{C}_{00} & \tilde{C}_{01} & \tilde{C}_{02}  & 0 & 0 & 0 \\
\tilde{C}_{01} & \tilde{C}_{00} & \tilde{C}_{02}  & 0 & 0 & 0 \\
\tilde{C}_{02} & \tilde{C}_{02} & \tilde{C}_{22}  & 0 & 0 & 0 \\
0 & 0 & 0 & \tilde{C}_{33} & 0 & 0 \\
0 & 0 & 0 & 0 & \tilde{C}_{33} & 0 \\
0 & 0 & 0 & 0 & 0 & \tilde{C}_{55}
\end{array} \right).
\end{align} 
We note that here the indices run from 0 to 5, not from 1 to 6.

We then turn to nonlinear elasticity.
The infinitesimal group generators corresponding to Eq.~\eqref{eq:linear_strain} are
$\{\tilde{\boldsymbol{\lambda}}_i\}$ for $i=0,\ldots,5$,
three of which are defined componentwise via 
$[\tilde{\boldsymbol{\lambda}}_{i-1}]_{lm} = \delta_{il}\delta_{im}$ for $i,l,m = 1,2,3$,
and the others by $\tilde{\boldsymbol{\lambda}}_i = \boldsymbol{\lambda}_i$ for $i=3,4,5$.
However, such a choice of group generators is 
not compatible with the symmetry of systems under the incompressibility constraint.
Therefore, we should introduce a transformation, which allows us to switch
to the generators of Eq.~\eqref{eq:SL3_generators}.
Since a set of infinitesimal group generators is a basis of a vector space,
namely Lie algebra, we can find the form of generalized elastic constants 
corresponding to Eq.~\eqref{eq:SL3_generators} via a linear transformation.
Specifically, a unitary transformation 
\begin{align}
\underline{\underline{U}} 
	= 	\left( \begin{array}{cccccc}  
			\frac{1}{\sqrt{3}} & \frac{1}{\sqrt{3}} & \frac{1}{\sqrt{3}} & 0 & 0 & 0 \\ 
	  	\frac{1}{\sqrt{2}} & -\frac{1}{\sqrt{2}} & 0 & 0 & 0 & 0 \\ 
	  \frac{1}{\sqrt{6}} & \frac{1}{\sqrt{6}} & -\frac{\sqrt{2}}{\sqrt{3}} & 0 & 0 & 0  \\\
	  0 & 0 & 0 & 1 & 0 & 0 \\
	  0 & 0 & 0 & 0 & 1 & 0 \\
	  0 & 0 & 0 & 0 & 0 & 1
	  \end{array} \right) 
\end{align}
connects, via $\underline{\epsilon} = \underline{\underline{U}}\cdot \tilde{\underline{\epsilon}}$,
the deformation vector 
$\tilde{\underline{\epsilon}} = 1/\sqrt{2} (\tilde{\epsilon}_0, \ldots, \tilde{\epsilon}_5)$ in Mandel notation,
corresponding to $\{\tilde{\boldsymbol{\lambda}}_{i}\}$,
to $\underline{\epsilon} = (\epsilon_0, \ldots, \epsilon_5)$, corresponding to 
the set of group generators consisting of $\{\boldsymbol{\lambda}_i\}$ for $i=1,\ldots,5$, 
and $\boldsymbol{\lambda}_0 \equiv \sqrt{2/3} \mathbf{I}$, 
where $\mathbf{I}$ is the $3\times 3$ identity matrix,
such that $\sum_i \epsilon_i \boldsymbol{\lambda}_i = \sum_i \tilde{\epsilon}_i \tilde{\boldsymbol{\lambda}}_i$.
Subsequently, the stiffness tensor $\dunder{C}_{\rm \,com}$ for compressible systems 
can be computed from 
$\dunder{C}_{\rm \,com} = \underline{\underline{U}} \cdot \dunder{C}_{\rm Mandel} \cdot \underline{\underline{U}}^T$,
as is obvious from linear algebra.
Within this representation, all deformations involving a volume change are associated with
the generator $\boldsymbol{\lambda}_0$. 
Therefore, under the incompressibility constraint, the components in $\dunder{C}_{\rm \,com}$ associated with 
$\boldsymbol{\lambda}_0$ become irrelevant.
Furthermore, predeformations also give rise to additional terms that are absent
in linear elasticity, as demonstrated in Ref.~\onlinecite{Goh2022}.
In our case, as only a predeformation in volume is involved, 
such nonlinear contributions are all diagonal and associated with the generalized pressure
\begin{align} \label{eq:pressure}
p = -\frac{1}{\sqrt{6}}\frac{\partial \mathcal{F}}{\partial \epsilon_0},
\end{align}
which only makes sense if a volume change is allowed.
Finally, we conclude that the stiffness tensor of incompressible systems takes the form of
\begin{align} \label{eq:stiff_incom_Tetra}
\dunder{C}_{\rm \,in}^{\rm Tetr} = \left( \begin{array}{ccccc}
		C_{11} & 0 & 0 & 0 & 0 \\  
		0 & C_{22} & 0 & 0 & 0 \\  
		0 & 0 & C_{33} & 0 & 0 \\  
		0 & 0 & 0 & C_{33} & 0 \\  
		0 & 0 & 0 & 0 & C_{55}  
\end{array} \right),
\end{align}
whose components are defined by Eq.~\eqref{eq:generalized_C_definition}.

Second, if the $(1,1,1)$-direction of the lattice is oriented 
along the $z$-axis, we have a rhombohedral lattice (RI Laue group),
the stiffness tensor of which, again in Mandel notation, 
is given as~\cite{Brugger1965, Clayton2010} 
\begin{align}
&\dunder{C}_{\rm Mandel}^{\rm Rhomb} \nonumber \\
		&= \left( \begin{array}{cccccc}
\tilde{C}_{00} & \tilde{C}_{01} & \tilde{C}_{02} & \sqrt{2} \tilde{C}_{03} & 0 & 0 \\
\tilde{C}_{01} & \tilde{C}_{00} & \tilde{C}_{02} & -\sqrt{2} \tilde{C}_{03} & 0 & 0 \\
\tilde{C}_{02} & \tilde{C}_{02} & \tilde{C}_{22} & 0 & 0 & 0 \\
\sqrt{2} \tilde{C}_{03} & -\sqrt{2} \tilde{C}_{03} & 0 & 2\tilde{C}_{33} & 0 & 0 \\
0 & 0 & 0 & 0 & 2\tilde{C}_{33} & 2\tilde{C}_{03} \\
0 & 0 & 0 & 0 & 2\tilde{C}_{03} & \tilde{C}_{00} - \tilde{C}_{01}
\end{array} \right).
\end{align} 
Then the stiffness tensor of the corresponding incompressible systems, within our notation, reads
\begin{align} \label{eq:stiff_rhomb}
\dunder{C}_{\rm \,in}^{\rm Rhomb} = \left( \begin{array}{ccccc}
		C_{11} & 0 & C_{13} & 0 & 0 \\  
		0 & C_{22} & 0 & 0 & 0 \\  
		C_{13} & 0 & C_{33} & 0 & 0 \\  
		0 & 0 & 0 & C_{33} & C_{13} \\  
		0 & 0 & 0 & C_{13} & C_{11}  
\end{array} \right).
\end{align}

As we demonstrate in Sec.~\ref{sec:result}, the macroscopic approach described here provides a precise 
and economic framework to investigate nonlinear elastic properties of incompressible anisotropic systems.
In particular, our choice of generators given by Eq.~\eqref{eq:SL3_generators} 
and the corresponding stiffness tensors given in Eqs.~\eqref{eq:stiff_incom_Tetra} and~\eqref{eq:stiff_rhomb}, respectively,  
determine all the necessary but only allowed deformations and elastic constants compatible with the underlying symmetry of the system and the imposed constraint.
Sticking to the linear strain tensors as given by Eq.~\eqref{eq:linear_strain}, instead of our nonlinear definition in Eq.~\eqref{eq:nonlinear_F}, 
may involve errors in the second order, which are relevant for elastic constants.
Indeed, the rotation coefficients $C_{66}$ and $C_{77}$ shown in Fig.~\ref{fig:rotation}(a) can become negative, 
if the volume conservation in the second order is not explicitly taken into account via Eqs.~\eqref{eq:DGT} and~\eqref{eq:SL3_generators}. 
Alternatively, one may consider the method of Lagrange multipliers, which is, however, technically demanding,
particularly in combination with the density functional calculation that we describe next.

\section{Density functional theory: Bridging scales}
\label{sec:DFT}
We now formulate a density functional theory (DFT) for the dipole-spring model
by approximating the free energy functional $\mathcal{F} [\rho(\mathbf{r})]$ 
where $\rho(\mathbf{r})$ denotes the one-body density field of the magnetic particles.
Together with the ideal gas term
\begin{align}
\mathcal{F}^{\rm id}[\rho (\mathbf{r})] = \beta^{-1} \int {\rm d}\mathbf{r} \, \rho (\mathbf{r}) [\ln{\{\Lambda^3 \rho (\mathbf{r})\}}-1],
\end{align}
where $\beta\equiv (k_\text{B}T)^{-1}$ is the inverse temperature,
the total free-energy functional subjected to minimization is given as
\begin{align}
\mathcal{F}[\rho (\mathbf{r})] = \mathcal{F}^{\rm id}[\rho (\mathbf{r})] + \mathcal{F}^{\rm ex}[\rho (\mathbf{r})],
\end{align}
where $\mathcal{F}^{\rm ex}[\rho (\mathbf{r})]$ denotes the excess functional describing the interparticle interactions~\eqref{eq_Hint}.
Following Ref.~\onlinecite{Oettel2010}, we employ the Picard iteration algorithm
\begin{align} \label{eq:minimization1}
\rho^{(i+1)}(\mathbf{r}) = \alpha \tilde{\rho}^{(i)} (\mathbf{r}) + (1-\alpha) \rho^{(i)} (\mathbf{r}),
\end{align}
with a mixing parameter $\alpha$ and 
\begin{align} \label{eq:minimization2}
\tilde{\rho}^{(i)} (\mathbf{r}) = \exp{\left( -\beta \frac{\delta \mathcal{F}^{\rm ex}}{\delta \rho(\mathbf{r})} +\beta \mu_i \right)},
\end{align}
where  
\begin{align}
\mu_i \equiv \frac{1}{V_{\rm cell}} \int_{\rm cell} {\rm d}\mathbf{r}\, \left\{ \ln{(\rho^{(i)} \Lambda^3)} -\frac{\delta \mathcal{F}^{\rm ex}}{\delta \rho^{(i)}} \right\},
\end{align}
which is updated in each iteration step to ensure that the total (average) number of particles is kept fixed.
Accordingly, for the verification of successful minimization, 
we use the relative chemical potential defined as
\begin{align}
\Delta \mu_{\rm rel} \equiv \frac{\mu_{i+1} - \mu_i}{\mu_{i+1}}.  
\end{align}
In this way, $\mathcal{F}$ is minimized for a prescribed value of the vacancy concentration $n_{\rm vac}$.
In principle, our model systems are defect-free, i.e., $n_{\rm vac} = 0$.
To accelerate and enhance the robustness of the minimization processes, however,
we consider lattices with vacancy concentration of $n_{\rm vac} = 0.001 \pm 10^{-6}$.  

Regarding the geometry of the calculation box,
we use the primitive unit cell in our calculations, consisting of only one
particle,
instead of the cubic unit cell of the FCC lattice,
consisting of four particles, usually adopted in DFT studies of freezing. 
Accordingly, we use periodic boundary conditions in the directions of three primitive vectors.
Both the primitive and reciprocal lattice vectors of undeformed and deformed systems 
are summarized in Appendix~\ref{sec:reciprocal}.
With this geometry, we are able to minimize our free-energy functional more precisely,
($\Delta \mu_{\rm rel} < 10^{-15}$ in most cases),
compared to the method using the cubic unit cell ($\Delta \mu_{\rm rel} \approx 10^{-8}$ for the tested cases).

Now we turn to the excess functional $\mathcal{F}^{\rm ex}[\rho(\mathbf{r})]$,
which is given as a sum of three functionals corresponding to the steric repulsion, 
magnetic dipole-dipole interaction, and harmonic spring potential.
First, for the hard-core repulsion, 
we use the White-Bear II (WB-II) functional~\cite{HansenGoos2006} 
with the Tarazona tensors~\cite{Tarazona2000},
which is one of the most precise versions among the fundamental measure theory 
for hard spheres~\cite{Roth2010}.
Then, for the elastic and magnetic dipole-dipole interactions,
we intend to adopt the simple mean-field functional in the form of
\begin{align} \label{eq:MF_functional}
\mathcal{F}_{\rm MF} [\rho (\mathbf{r})]\equiv \frac{1}{2}\int {\rm d}\mathbf{r}\int {\rm d}\mathbf{r}'\, 
	\rho (\mathbf{r}) \, u (\mathbf{r} - \mathbf{r}') \, \rho (\mathbf{r}'), 
\end{align}
where $u (\mathbf{r})$ is an appropriate pair potential.
However, the practical evaluation 
of the above functionals is not straightforward.
In what follows, we describe how to construct the Fourier transforms of the elastic and magnetic energies,
which allow us to perform DFT calculations in Fourier space. 

\subsection{Magnetic dipolar interaction}
\label{sec:magnetic}
As discussed, there are two important properties
inherent in the magnetic dipole-dipole interaction, Eq.~\eqref{eq:u_m}, in three dimensions.
It is long-range and anisotropic~\cite{Klapp2005},
which has to be taken into account in DFT calculations. 

When we switch $m$ to $m \neq 0$, the systems elongate or contract, and so does the unit cell.
Then the side lengths of the cubic unit cell are no longer the same,
but satisfy the relation $a_x = a_y \neq a_z$
where $a_x$, $a_y$, and $a_z$ denote the side lengths in the $x$-, $y$-, and $z$-direction, respectively.
Here, we define the aspect ratio as $R_{\rm asp} \equiv a_z/ a_x$.
We note that $R_{\rm asp}$ characterizes the 
deformation of the internal lattice structure. 

Now, we address the long-range nature of the magnetic dipole-dipole interaction in three dimensions.
The difficulty arises from the fact 
that the interaction energy, i.e., the integral of $u_{\rm m}$, diverges at both short and long distances.
In our DFT calculation, this issue can be resolved rather easily.
On the one hand, the steric repulsion hinders particles from approaching closer than their diameter and therefore prevents the divergence at short distances.
On the other hand, as the DFT calculation is performed in Fourier space, 
the divergence at long distance can be handled directly as follows.
While, for $\mathbf{k}\neq 0$ Fourier modes, the Fourier transform of the dipole-dipole interaction 
can be obtained with the aid of the orthogonality of the spherical harmonics $Y_l^m$,
the $\mathbf{k}=0$ mode, which dictates the long-range divergence,
indeed depends on the shape of the whole material body (see Appendix~\ref{sec:FT_magnetic} for more details).
With such a shape dependent mode, 
which is related to the demagnetizing factor in continuum theory~\cite{Ivaneyko2014},
we are able to capture the long-ranged nature of the magnetic interaction.
In general, we may consider a system with the initially spheroidal shape (at $m = 0$)
of the shape parameter $R_{\rm sh} \equiv R_z /R_x$, where $R_x=R_y$ and $R_z$ are the lengths of the semiaxes along the $x$-, $y$- and $z$-axis, respectively.
In contrast to $R_{\rm asp}$, 
here $R_{\rm sh}$ indicates the aspect ratio of the whole material. 
As we turn on the magnetic interaction applying a magnetic field,
the \emph{initial} aspect ratio of the whole system shape further changes 
to $R_{\rm asp} R_{\rm sh}$ due to magnetostriction associated with a change 
in internal lattice structure.  

\subsection{Elastic energy}
\label{sec:elasticenergy}
While the magnetic particles are strictly labeled due to fixation by the surrounding polymer matrix,
namely the elastic interaction term given as Eq.~\eqref{eq_Hel}, 
the conventional machinery of DFT calculation assumes the indistiguishability of particles, i.e., as if the potential $u_{\rm el}$ was acting equally between all pairs of particles throughout the system.
To nevertheless enable DFT calculations, a mapping of the harmonic spring potential onto a pseudospring potential $u_{\rm pseudo}$
has been proposed in Ref.~\onlinecite{Cremer2017}. 
There, only nearest-neighbor pairs of the resulting configuration 
are within the range of $u_{\rm pseudo}$ and thus elastically coupled to each other,
as in the original system based on the harmonic springs, see Eq.~\eqref{eq_Hel}.
In the present study, the mapping is extended to three dimensions.
Notably, in two and three spatial dimensions,
the success of applying the finite-ranged pseudospring potential 
is connected to the particle arrangement arising from
a freezing transition~\cite{Goh2019}, which has been extensively investigated
within density functional approaches~\cite{Ramakrishnan1979,Curtin1985,Denton1989,Rosenfeld1989,Ohnesorge1993}. 
Specifically, we consider the pseudospring potential
\begin{align} \label{eq:pseudo}
u_{\rm pseudo} (r)=\begin{cases}
	\frac{1}{2}k_{\rm el} (r-a_{\rm el})^2 - u_0, & 
 r < R_c  \\
	0, & {\rm otherwise}.
\end{cases}
\end{align}
In this expression, $R_c$ and $u_0$ denote the cut-off length and the offset for the pseudospring potential, respectively.
The cutoff length $R_c$ is determined from corresponding Monte-Carlo simulations 
as the distance at which the pair correlation function $g(r)$ is minimized, 
which turns out to be $R_c = 1.21 a_{\rm el}$.
The value of $u_0$ is determined within the DFT calculations 
so as to match the vacancy concentration of the resultant lattice 
with the prescribed value of $n_{\rm vac} = 0.001$.

The obtained $u_{\rm pseudo}$ instead of $u_{\rm el}$ is inserted into Eq.~\eqref{eq:MF_functional} via $u$.
We refer to our previous study~\cite{Goh2019} for the detailed description and verification
of the mapping of the real onto the pseudospring potential.
Moreover, due to the anisotropy of dipolar interactions as discussed in Sec.~\ref{sec:magnetic},
corresponding lattice structures may become anisotropic as well.
Such anisotropy then should also be taken into account when we construct the pseudospring potential.
In practice, we then cut the spring potential at the surface of the spheroid
with the aspect ratio $R_{\rm asp}$, 
instead of at the surface of the sphere with the radius $R_c$ as in Eq.~\eqref{eq:pseudo}.
In other words, the cutting is direction-dependent.
The resultant Fourier components are presented in Appendix~\ref{sec:FT_pseudo} explicitly.

\section{Mechanical properties}
\label{sec:result}
From now on, we measure lengths and energies in units of the rest length $a_{\rm el}$ and the thermal energy $k_\text{B} T$, respectively.
Accordingly, the magnitude $m$ of the magnetic moment and the spring constant $k_{\rm el}$ are measured 
in units of $m_0 \equiv \sqrt{k_\text{B} T a_{\rm el}^3 /\mu_0}$ and $k_\text{B} T/a_{\rm el}^2$, respectively.
We consider systems with elastic constant $k_{\rm el}=100$ and shape parameter $R_{\rm sh} = 1$,
and investigate the effects of magnetization on the mechanical properties,
by varying the magnitude of the magnetic moment $m$. 
The two models, described in Sec.~\ref{sec:model}, give identical results as long as no rotations are considered,
while only the paramagnetic Model II has a unique reference state with respect to rotations. 

One can also probe steric effects by varying the volume packing fraction $\eta$.
Naively speaking, while steric repulsion should affect the bulk modulus of a system,
how and to what extent it would affect the mechanical properties under each specific deformation is still unclear.
Moreover, there might also appear numerical artifacts due to several approximations employed.
Therefore, leaving systematic investigations for further studies,
we demonstrate that our method is valid for a reasonable range of $\eta$ 
by employing two representative values of $\eta = 0.1$ and 0.3, 
which are relatively small when compared to
the coexisting fluid (crystal) packing fractions 0.495 (0.544)~\cite{Oettel2010}
for the WB-II functional used in this study.
We note that the pseudospring potential suffices to stabilize the FCC crystal 
for $\eta = 0$ within our model.  
Indeed, steric repulsion does not play a dominant role for these low packing fractions, 
as one may confirm from Fig.~\ref{fig:magnetostriction} in Sec.~\ref{sec:magnetostriction}, 
as well as from Figs.~\ref{fig:elastic} and~\ref{fig:rotation} in Sec.~\ref{sec:elastic}.

\subsection{Magnetostriction}
\label{sec:magnetostriction}

\begin{figure}
\includegraphics[width=8.5cm]{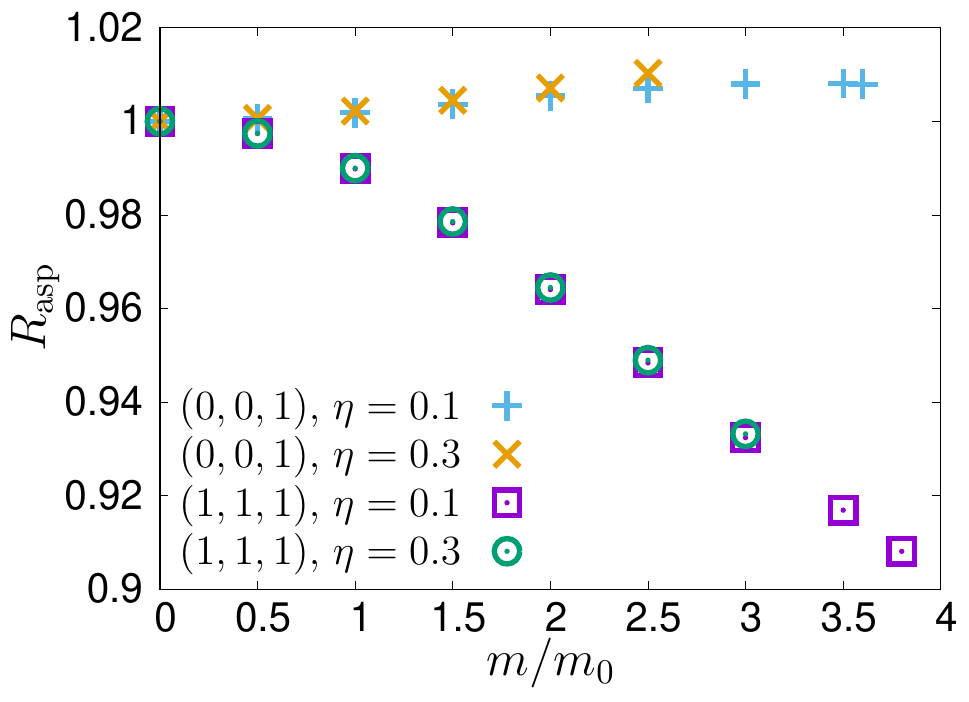} 
\caption{\label{fig:magnetostriction}
The aspect ratio $R_{\rm asp}$ of the system is presented as a function of $m$.
Converse magnetostriction effects are observed, depending on the orientation of the lattice.}
\end{figure}

As a first step, we determine the reference equilibrium state of the undeformed system for a given magnetization $m$. 
Technically, we first determine the value of $u_0$ for which 
the vacancy concentration of system becomes equal to the prescribed value within the margin of tolerated error.
Then, varying $R_{\rm asp}$ while fixing $u_0$,
we find the value of the aspect ratio $R_{\rm asp}$ at which the free energy functional is minimized.
The resultant values of $R_{\rm asp}$ are shown in Fig.~\ref{fig:magnetostriction}.

The most prominent feature here is that 
the magnetostriction effects of the $(0,0,1)$- and $(1,1,1)$-orientations are opposite to each other.
In line with the results reported in Ref.~\onlinecite{Fischer2019}, our systems elongate when 
the dipole moments are directed along the $(0,0,1)$-orientation,
while a contraction along the direction of the dipole moments is observed in the $(1,1,1)$-case,
confirming that the internal configuration of magnetic particles is a decisive factor of the magnetostriction effect.
In addition, we also note that the magnetostriction effect can be reversed,
if large values of the shape parameter ($R_{\rm sh} \gtrsim 2$) are used 
in the case of the $(0,0,1)$-orientation (results not shown). 
Such shape-dependence is a trivial consequence of the long-range nature of the dipolar interaction.

\subsection{Elastic coefficients}
\label{sec:elastic}

\begin{figure*}
\includegraphics[width=17cm]{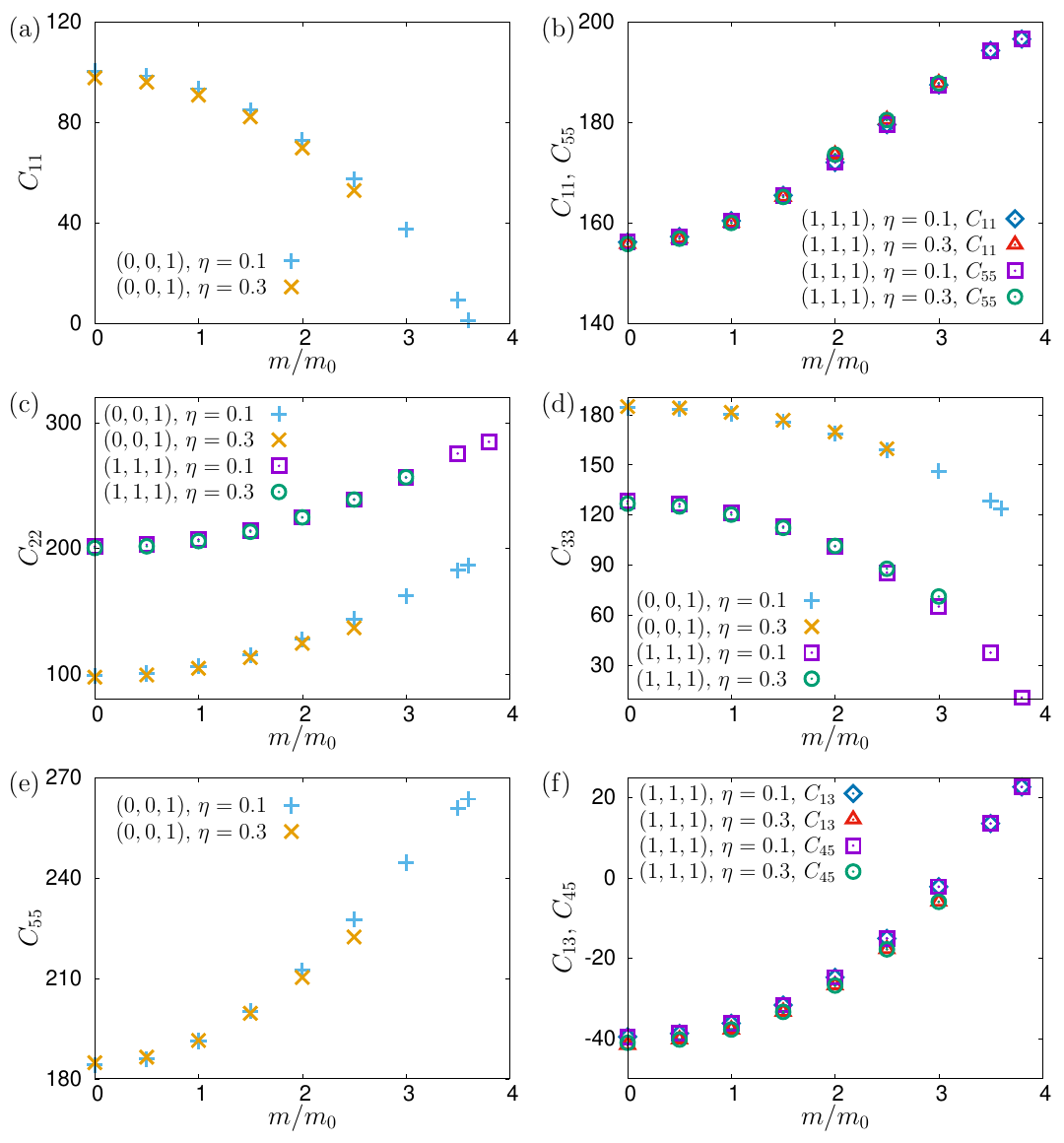} 
\caption{\label{fig:elastic}
Elastic coefficients are presented as functions of $m$. The values of 
(a) $C_{11}$ for the $(0,0,1)$-orientation, 
(b) $C_{11}$ and $C_{55}$ for the $(1,1,1)$-orientation, 
(c) $C_{22}$ for the $(0,0,1)$- and $(1,1,1)$-orientations, 
(d) $C_{33}$ for the $(0,0,1)$- and $(1,1,1)$-orientations,
(e) $C_{55}$ for the $(0,0,1)$-orientation, and 
(f) $C_{13}$ and $C_{45}$ for the $(1,1,1)$-orientation 
are shown. For values of $m$ larger than those presented in this figure, we were not able to find stable equilibrium configurations.
}
\end{figure*}

Now we determine the elastic constants $C_{ij}$ for $i, j=1, \ldots, 5$, defined in Eq.~\eqref{eq:generalized_C_definition}, from our DFT,
explicitly deforming the primitive unit cell.
Specifically, we numerically calculate the derivatives through finite differences and
obtain the diagonal terms of the stiffness tensor from
\begin{align} \label{eq:diagonal_numerical}
C_{ii} = \left(\frac{\partial^2 \mathcal{F}}{{\partial \epsilon_i}^2}\right) \approx \frac{\mathcal{F}(\epsilon_i)+\mathcal{F}(-\epsilon_i)-2\mathcal{F}(0)}{\epsilon_i^2},
\end{align}
while the offdiagonal terms can be calculated as
\begin{align} \label{eq:offdiagonal_numerical}
&C_{ij} = \left( \frac{\partial^2 \mathcal{F}}{\partial \epsilon_i \partial \epsilon_j}\right) \nn \\
&\approx \frac{\mathcal{F}(\epsilon_i, \epsilon_j) + \mathcal{F}(-\epsilon_i, -\epsilon_j)
 -\mathcal{F}(-\epsilon_i, \epsilon_j) - \mathcal{F}(\epsilon_i, -\epsilon_j)}{4\epsilon_i \epsilon_j}.
\end{align}
In most of the cases, we use
$\epsilon_1=\epsilon_2/\sqrt{3} = \epsilon_3 = \cdots = \epsilon_7 = 0.0001$,
except for the cases of the $(1,1,1)$-orientation with $\eta = 0.3$ and $m \leq 2.5$, 
in which the functional can be minimized up to the values of $\Delta \mu_{\rm rel}$ between $10^{-4}$ and $10^{-8}$ at most.
There, we use $\epsilon_1=\epsilon_2/\sqrt{3} = \epsilon_3 = \cdots = \epsilon_7 = 0.0001$ 
or $0.001$ to obtain consistent results.
Before proceeding to the results in Fig.~\ref{fig:elastic}, we  recall from Sec.~\ref{sec:elasticity} that some coefficients vanish and others are equal.
Specifically, we confirm in Figs.~\ref{fig:elastic}(b) and (f) for rhombohedral lattices that $C_{11} = C_{55}$ and $C_{13} = C_{45}$, respectively,
 in accordance with Eq.~\eqref{eq:stiff_rhomb}. 

First, we take a closer look at the 
elastic constant $C_{55}$, corresponding to shear deformations in the $xy$-plane,
and $C_{11}$, corresponding to stretches (compressions) along the $x$-axis combined with compressions (stretches) along the $y$-axis. 
$C_{11}$ can also be regarded as a shear modulus, but corresponding to 
shear deformations with orientations different from those for $C_{55}$. 
In most cases, the dipolar interaction, which is repulsive in the plane perpendicular to the dipole moment, 
causes an increase of the elastic constants.
Specifically, as shown in Fig.~\ref{fig:elastic}(b),
values of both $C_{11}=C_{55}$ increase as $m$ increases in the $(1,1,1)$-case,
while, in the $(0,0,1)$-case, only $C_{55}$ is an increasing function of $m$, as shown in Fig.~\ref{fig:elastic}(e). In sharp contrast, 
$C_{11}$ in the $(0,0,1)$-case is a decreasing function of $m$, as shown in Fig.~\ref{fig:elastic}(a).
Furthermore, as $m$ increases further, it drops towards zero,
indicating instability of the tetragonal lattices.
We notice here that hexagonal configurations can be obtained
eventually by squeezing the tetragonal lattice in the $xy$-plane,
if the whole lattice is projected on the $xy$-plane.
In other words, as $m$ increases, there might arise a growing tendency 
to match the lattice to the underlying symmetry of the magnetic dipole-dipole interaction, 
which prefers the hexagonal lattice over the tetragonal lattice in the plane perpendicular to the dipole moment.
Therefore, we conclude that such a softening effect 
correlates with a rearrangement of the magnetic particles 
in the plane perpendicular to $\mathbf{m}$.

\begin{figure}[h!]
\includegraphics[width=8.5cm]{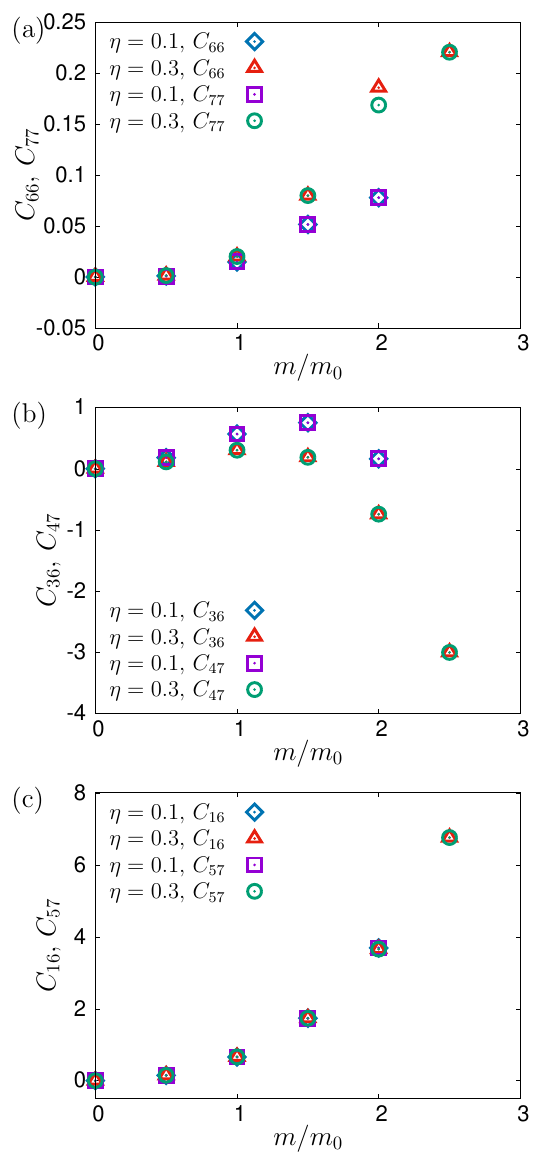} 
\caption{\label{fig:rotation}
Rotation coefficients obtained from the $(1,1,1)$-orientation. 
In (a), the coefficients $C_{66}$ and $C_{77}$, respectively, corresponding 
solely to the rotations in the $yz$- and $xz$-planes are presented, 
whereas the off-diagonal coefficients $C_{36}$ and $C_{47}$ as well as $C_{16}$ and $C_{57}$ 
are depicted in (b) and (c), respectively.}
\end{figure}

Next we turn to the elastic constants of
$C_{22}$, corresponding to stretches (compressions) in the $xy$-plane combined with compressions (stretches) along the $z$-axis of twice the magnitude, 
and $C_{33}$, corresponding to shear deformation in the $yz$-plane 
(or equivalently $C_{44}$, corresponding to shear deformations in the $xz$-plane).
All of them involve deformations in the $z$-direction.
In both the $(0,0,1)$- and $(1,1,1)$-orientations, $C_{22}$ is an increasing function of $m$ 
[Fig.~\ref{fig:elastic}(c)], indicating hardening of the materials.
Since there is no significant 
difference between the systems of $\eta=0.1$ and $0.3$, 
the phenomenon of hardening observed here has a purely elastic origin.
Simultaneously, $C_{33}$ is always a decreasing function of $m$ [Fig.~\ref{fig:elastic}(d)].
Moreover, at large $m$, the rhombohedral lattice becomes unstable as well, 
with the values of $C_{33}$ dropping towards zero.
Such instabilities at large $m$ and the decrease of $C_{33}$ in general
may originate from the tendency towards pair formation~\cite{Pessot2018, Goh2018} or similarly 
from the typical chain-like aggregates forming under strong dipolar interactions~\cite{Menzel2014,Allahyarov2015,Jaeger2022}.
Indeed, we observe 
a shift of the energetic minimum in the landscape of two-body interaction energy from separated to touching configurations
occurs between $m=2.5$ and $3.0$ in the $(1,1,1)$-case (not shown).
This seems to confirm that the instability is the consequence of the formation of touching pairs. 
In the case of the $(0,0,1)$-orientation, the drop towards zero in $C_{11}$ occurs
in advance of that in $C_{33}$, compare Figs.~\ref{fig:elastic}(a) and (d), 
indicating that rearrangement in the $xy$-plane is preferred over rearrangement in the $z$-direction. 

Lastly, the values of $C_{13}$ and $C_{45}$ in the $(1,1,1)$-case are presented  
in Fig.~\ref{fig:elastic}(f).
Overall they exhibit a similar behavior, increasing from
 negative values for small $m$ to positive ones for large $m$. 
We note, however, that these constants reflect a rather specific symmetry inherent in the lattice,
and therefore, may not reflect the situation of real magnetic gels.

\subsection{Rotation coefficients}
\label{sec:rotation}

Finally, we investigate the rotation coefficients, which are relevant only in Model II.
Alike the elastic constants,
the rotation coefficients are calculated from 
Eqs.~\eqref{eq:diagonal_numerical} and \eqref{eq:offdiagonal_numerical}.
As the $(0,0,1)$-orientation turns out to be unstable with respect to rotations
in $xz$- and $yz$-planes,
we only analyze the results for the $(1,1,1)$-orientation.

First, the coupling of the model systems to the applied magnetic field is captured
by the rotation coefficients $C_{66}$ and $C_{77}$, 
corresponding to rotations in the $yz$- and $xz$-plane, respectively.
As shown in Fig.~\ref{fig:rotation}(a),
the values of $C_{66}$ and $C_{77}$ increase as $m$ increases,
indicating an enhanced resistance to the rotations.

As shown in Fig.~\ref{fig:rotation}(b), the mixed coefficients of $C_{36}$ and $C_{47}$,
corresponding to mixed shear deformations and rotations in the $yz$- and $xz$-plane, respectively,
first exhibit an increase as a function of $m$ for small values of $m$.
Then, the increasing trend is reversed for large $m$. 
We note that, in Ref.~\onlinecite{Menzel2014}, where chain-like aggregates are assumed, 
only a decreasing tendency in the form of $-m^2$ has been predicted for $D_2$,
which is equivalent to $C_{36}$ and $C_{47}$ in the present study.
Presumably, as already mentioned for magnetostrictive effects in Sec.~\ref{sec:magnetostriction}, 
different behaviors may be due to the internal configuration of the magnetic particles. 
We also note that the values of $C_{66}$ and $C_{77}$ are approximately $10^3$ times smaller 
than those of $C_{33}$ and $C_{44}$. In Ref.~\onlinecite{Menzel2014},
the rotation coefficient $D_1$ (equivalent to $C_{66}$ and $C_{77}$ in the present study)
is even larger than $\Delta c_{5}$ ($C_{33}$ and $C_{44}$ in the present study).
Again, this may be caused by the different internal structure,
which is chain-like in Ref.~\onlinecite{Menzel2014}.

Lastly, the additional mixed coefficients $C_{16}$ and $C_{57}$ increase monotonically, as shown in
Fig.~\ref{fig:rotation}(c),
which seems to be a simple consequence of enhancement of both hardening in the $xy$-plane 
($C_{11}$ and $C_{55}$)
and resistance to rotations in the $xz$- or $yz$-plane ($C_{66}$ and $C_{77}$).

\section{Conclusion}
\label{sec:discussion}
So far we have constructed and evaluated a DFT for three-dimensional dipole-spring models,
which bridges from the discretized mesoscopic model to a macroscopic elasticity theory
of magnetic gels.
Based on the scale-bridging description, 
we have determined the elastic and rotational material coefficients. 
They depend on the mesoscopic configuration of the magnetic particles.
Notably, we have observed softening responses to magnetization
both in the external field direction and in the plane perpendicular to the external field,
which indicates a tendency towards an instability.
We have proposed that such behaviors imply changes in overall symmetry,
accompanied by rearrangement of magnetic particles.
Such rearrangements might be decomposed into 
the formation of a hexagonal-like arrangement in the plane perpendicular to the magnetic field 
and pair formation along the magnetic field direction.
To verify our conclusion, the decreasing behavior of $C_{11}$ needs to be tested experimentally.
Notably, in a previous study, where random configurations for magnetic particles are assumed 
[see Fig.~14(b) of Ref.~\onlinecite{Pessot2016}], 
a decrease of the shear modulus has been observed,
suggesting that the idea of rearrangement may also be valid 
for real magnetic gels with disordered configurations.

Conversely, one could equally well think about synthesizing
a sample with the regular arrangement adopted in this study.
In particular, the prescribed FCC-based connectivity 
shows certain characteristics as explained above.
For instance, 6 among 12 nearest neighboring particles are located in the same plane
perpendicular to the magnetization
in the case of $(1,1,1)$-orientation, and thereby, the repulsive interaction in the plane seems to dominate
the response of the magnetic particles. 
This leads to the contraction along the magnetization direction.
We note that there have been attempts to synthesize thin ferrogel films~\cite{Queralto2010}.
Since in planar configurations, magnetic particles form hexagonal arrangements
in the plane perpendicular to the external magnetic field~\cite{vanTeeffelen2008},
it would be possible to obtain ferrogel films with a hexagonal configuration in such a way.
Then, by stacking two-dimensional layers,
a magnetic gel with a three-dimensional hexagonal structure might be fabricated experimentally.
Our results of the $(1,1,1)$-case may then provide an insight into such systems.

At the same time, regarding future work on our theory, one important direction is to address systems with random configurations.
An important additional ingredient to model the heterogeneity inherent in real samples
is polydispersity of the magnetic particles~\cite{Fischer2020}.
The idea of the replica DFT~\cite{Reich2004, Schmidt2005} 
might be used to address directly disordered configurations.
Lastly, dynamical density functional theory~\cite{Marconi1999,Schmidt2013,teVrugt2020} 
should provide a route to investigate the dynamics of the systems.

\section*{Acknowledgment}
This work was supported by funding from the Deutsche Forschungsgemeinschaft (DFG) through the SPP 1681, 
Grant No. ME 3571/3-3 (A.M.M.) and Grant No. LO 418/16-3 (H.L.).
A.M.M is presently supported by the DFG through Heisenberg Grant No. ME 3571/4-1.

%
%
%

\appendix

\begin{widetext}
\section{Reciprocal lattices}
\label{sec:reciprocal}

For the $(0,0,1)$-, and $(1,1,1)$-orientations of the FCC lattice,
the primitive vectors read
\begin{align} \label{eq:primitive_001}
\mathbf{a}_1 = \frac{a}{2}(\hat{x}+\hat{y}), \quad
\mathbf{a}_2 = \frac{a}{2}(\hat{y}+\hat{z}), \quad
\mathbf{a}_3 = \frac{a}{2}(\hat{z}+\hat{x}),
\end{align}
and
\begin{align} \label{eq:primitive_111}
\mathbf{a}_1 = \frac{a}{2}\left(\frac{2}{\sqrt{6}}\hat{y} +\frac{2}{\sqrt{3}}\hat{z}\right), \quad
\mathbf{a}_2 = \frac{a}{2}\left(-\frac{1}{\sqrt{2}}\hat{x} -\frac{1}{\sqrt{6}}\hat{y}+\frac{2}{\sqrt{3}}\hat{z}\right), \quad
\mathbf{a}_3 = \frac{a}{2}\left(\frac{1}{\sqrt{2}}\hat{x}-\frac{1}{\sqrt{6}}\hat{y}+\frac{2}{\sqrt{3}} \hat{z}\right),
\end{align}
respectively.
Here, $a = \sqrt{2} a_{\rm el}$ denotes the side length of the cubic unit cell.

In practice, the DFT calculations are performed with the reciprocal lattice vectors in Fourier space.
For the $(0,0,1)$-orientation, the reciprocal vectors read 
\begin{align}
\mathbf{b}_1 = \frac{2\pi}{a} \left( 1, 1, -1 \right), \quad
\mathbf{b}_2 = \frac{2\pi}{a} \left( -1, 1, 1 \right), \quad
\mathbf{b}_3 = \frac{2\pi}{a} \left( 1, -1, 1 \right),
\end{align}
while for the $(1,1,1)$-orientation, we obtain
\begin{align}
\mathbf{b}_1 = \frac{2\pi}{a} \left( 0, \frac{4}{\sqrt{6}}, \frac{1}{\sqrt{3}} \right), \quad
\mathbf{b}_2 = \frac{2\pi}{a} \left( -{\sqrt{2}}, -\frac{2}{\sqrt{6}}, \frac{1}{\sqrt{3}} \right), \quad
\mathbf{b}_3 = \frac{2\pi}{a} \left( {\sqrt{2}}, -\frac{2}{\sqrt{6}}, \frac{1}{\sqrt{3}} \right).
\end{align}

Under deformation, the reciprocal vectors are transformed accordingly.
We expand the reciprocal vectors of deformed lattices with respect to $\{\epsilon_i\}$  
to compute the corresponding reciprocal lattice vectors in the form
\begin{align}
\mathbf{b}_1^{\rm deformed} = \mathbf{b}_1 + \frac{2\pi}{a}\Delta \mathbf{b}_1, \quad 
\mathbf{b}_2^{\rm deformed} = \mathbf{b}_2 + \frac{2\pi}{a}\Delta \mathbf{b}_2, \quad 
\mathbf{b}_3^{\rm deformed} = \mathbf{b}_3 + \frac{2\pi}{a}\Delta \mathbf{b}_3. 
\end{align}
The correction terms $\Delta \mathbf{b}_1$, $\Delta \mathbf{b}_2$, and $\Delta \mathbf{b}_3$ 
for the $(0,0,1)$- and $(1,1,1)$-cases are given 
in Tables~\ref{table:reciprocal_deformed_001} and \ref{table:reciprocal_deformed_111}, respectively, 
which are sufficient for the pure deformations that do not involve mixed terms, i.e., $\epsilon_i \epsilon_j$ for $i\neq j$.
When more than two different types of deformations are applied, 
Eq.~\eqref{eq:DGT} still provides a correct formulation.
However, such mixed terms are irrelevant for our incompressible systems
because second-order corrections only enter via the diagonal terms in the stiffness tensor,
as we describe in Sec.~\ref{sec:irreducible} (see Ref.~\onlinecite{Goh2022} for details).
Therefore, for the calculation of off-diagonal components in the stiffness tensors,
we simply add the second-order corrections from two different types of pure deformations. 

\renewcommand{\arraystretch}{1.8}
\begin{table*}
\begin{tabular}{c|ccc}
\hline
\hline
$\epsilon$ & $\Delta \mathbf{b}_1$ & $\Delta \mathbf{b}_2$ & $\Delta \mathbf{b}_3$ \\
\hline
$\epsilon_1$ &
$\left( -\epsilon +\frac{1}{2}\epsilon^2, \epsilon +\frac{1}{2}\epsilon^2, 0 \right)$ & 
$\left( \epsilon -\frac{1}{2}\epsilon^2, \epsilon +\frac{1}{2}\epsilon^2, 0\right)$ & 
$\left( -\epsilon+\frac{1}{2}\epsilon^2, -\epsilon-\frac{1}{2}\epsilon^2, 0 \right)$ \\ 
$\epsilon_2$ &
$\left( -\frac{1}{\sqrt{3}}\epsilon +\frac{1}{6}\epsilon^2, -\frac{1}{\sqrt{3}}\epsilon+\frac{1}{6}\epsilon^2, -\frac{2}{\sqrt{3}}\epsilon-\frac{2}{3}\epsilon^2 \right)$ & 
$\left( \frac{1}{\sqrt{3}}\epsilon-\frac{1}{6}\epsilon^2, -\frac{1}{\sqrt{3}}\epsilon+\frac{1}{6}\epsilon^2, \frac{2}{\sqrt{3}}\epsilon+\frac{2}{3}\epsilon^2 \right)$ & 
$\left( -\frac{1}{\sqrt{3}}\epsilon+\frac{1}{6}\epsilon^2, \frac{1}{\sqrt{3}}\epsilon-\frac{1}{6}\epsilon^2, \frac{2}{\sqrt{3}}\epsilon+\frac{2}{3} \epsilon^2 \right)$ \\ 
$\epsilon_3$ &
$\left( 0, \epsilon +\frac{1}{2}\epsilon^2, -\epsilon-\frac{1}{2}\epsilon^2 \right)$ &
$\left( 0, -\epsilon +\frac{1}{2}\epsilon^2, -\epsilon +\frac{1}{2}\epsilon^2 \right)$ &
$\left( 0, -\epsilon -\frac{1}{2}\epsilon^2, \epsilon +\frac{1}{2}\epsilon^2 \right)$ \\
$\epsilon_4$ &
$\left( \epsilon+\frac{1}{2}\epsilon^2, 0, -\epsilon-\frac{1}{2}\epsilon^2 \right)$ & 
$\left( -\epsilon-\frac{1}{2}\epsilon^2, 0, \epsilon+\frac{1}{2}\epsilon^2 \right)$ & 
$\left( -\epsilon +\frac{1}{2}\epsilon^2, 0, -\epsilon+\frac{1}{2}\epsilon^2 \right)$ \\ 
$\epsilon_5$ &
$\left( -\epsilon +\frac{1}{2}\epsilon^2, -\epsilon+\frac{1}{2}\epsilon^2, 0 \right)$ & 
$\left( -\epsilon -\frac{1}{2}\epsilon^2, \epsilon+\frac{1}{2}\epsilon^2, 0 \right)$ & 
$\left( \epsilon+\frac{1}{2}\epsilon^2, -\epsilon-\frac{1}{2}\epsilon^2, 0 \right)$ \\ 
$\epsilon_6$ &
$\left(0, -\epsilon-\frac{1}{2}\epsilon^2, -\epsilon+\frac{1}{2}\epsilon^2 \right)$ &
$\left(0, \epsilon-\frac{1}{2}\epsilon^2, -\epsilon-\frac{1}{2}\epsilon^2 \right)$ &
$\left(0, \epsilon+\frac{1}{2}\epsilon^2, \epsilon-\frac{1}{2}\epsilon^2 \right)$ \\
$\epsilon_7$ &
$\left( -\epsilon-\frac{1}{2}\epsilon^2, 0, -\epsilon+\frac{1}{2}\epsilon^2 \right)$ & 
$\left( \epsilon+\frac{1}{2}\epsilon^2, 0, \epsilon-\frac{1}{2}\epsilon^2 \right)$ & 
$\left( \epsilon-\frac{1}{2}\epsilon^2, 0, -\epsilon-\frac{1}{2}\epsilon^2 \right)$ \\ 
\hline
\hline
\end{tabular}
\caption{\label{table:reciprocal_deformed_001} 
The reciprocal lattice vectors for the $(0,0,1)$-orientation.
As for the infinitesimal parameters $\{\epsilon_i\}$,
see Eqs.~\eqref{eq:DGT} and~\eqref{eq:SL3_generators} 
in which the deformation gradient tensor as well as the generators are defined.
}
\end{table*}

\renewcommand{\arraystretch}{1.8}
\begin{table*}
\begin{tabular}{c|ccc}
\hline
\hline
$\epsilon$ & $\Delta \mathbf{b}_1$ & $\Delta \mathbf{b}_2$ & $\Delta \mathbf{b}_3$ \\
\hline
$\epsilon_1$ &
$\left(0, \frac{4}{\sqrt{6}}\epsilon +\frac{2}{\sqrt{6}}\epsilon^2, 0 \right)$ &
$\left( \sqrt{2}\epsilon -\frac{\sqrt{2}}{2}\epsilon^2, -\frac{2}{\sqrt{6}}\epsilon -\frac{1}{\sqrt{6}}\epsilon^2, 0 \right)$ &
$\left( -\sqrt{2}\epsilon +\frac{\sqrt{2}}{2}\epsilon^2, -\frac{2}{\sqrt{6}}\epsilon -\frac{1}{\sqrt{6}}\epsilon^2, 0 \right)$ \\
$\epsilon_2$ & 
$\left( 0, -\frac{2\sqrt{2}}{3}\epsilon+\frac{\sqrt{6}}{9}\epsilon^2, \frac{2}{3} \epsilon+ \frac{2\sqrt{3}}{9}\epsilon^2 \right)$ &
$\left( \frac{\sqrt{6}}{3}\epsilon-\frac{\sqrt{2}}{6}\epsilon^2, \frac{\sqrt{2}}{3}\epsilon-\frac{\sqrt{6}}{18}\epsilon^2, \frac{2}{3}\epsilon+\frac{2\sqrt{3}}{9} \epsilon^2 \right)$ & 
$\left( -\frac{\sqrt{6}}{3}\epsilon+\frac{\sqrt{2}}{6}\epsilon^2, \frac{\sqrt{2}}{3}\epsilon-\frac{\sqrt{6}}{18}\epsilon^2, \frac{2}{3}\epsilon+\frac{2\sqrt{3}}{9}\epsilon^2 \right)$ \\
$\epsilon_3$ &
$\left( 0, -\frac{1}{\sqrt{3}}\epsilon+\frac{2}{\sqrt{6}}\epsilon^2, -\frac{4}{\sqrt{6}}\epsilon+\frac{\sqrt{3}}{6}\epsilon^2 \right)$ & 
$\left( 0, -\frac{1}{\sqrt{3}}\epsilon-\frac{1}{\sqrt{6}}\epsilon^2, +\frac{2}{\sqrt{6}}\epsilon+\frac{\sqrt{3}}{6}\epsilon^2 \right)$ & 
$\left( 0, -\frac{1}{\sqrt{3}}\epsilon-\frac{1}{\sqrt{6}}\epsilon^2, \frac{2}{\sqrt{6}}\epsilon +\frac{\sqrt{3}}{6}\epsilon^2 \right)$ \\ 
$\epsilon_4$ &
$\left( -\frac{1}{\sqrt{3}}\epsilon, 0, \frac{\sqrt{3}}{6}\epsilon^2 \right)$ &
$\left( -\frac{1}{\sqrt{3}}\epsilon-\frac{1}{\sqrt{2}}\epsilon^2, 0, \sqrt{2}\epsilon +\frac{\sqrt{3}}{6}\epsilon^2 \right)$ &
$\left( -\frac{1}{\sqrt{3}}\epsilon+\frac{1}{\sqrt{2}}\epsilon^2, 0, -\sqrt{2}\epsilon +\frac{\sqrt{3}}{6}\epsilon^2 \right)$ \\
$\epsilon_5$ & 
$\left( -\frac{4}{\sqrt{6}}\epsilon, \frac{2}{\sqrt{6}} \epsilon^2, 0 \right)$ &
$\left( \frac{2}{\sqrt{6}}\epsilon-\frac{1}{\sqrt{2}}\epsilon^2, \sqrt{2}\epsilon-\frac{1}{\sqrt{6}}\epsilon^2, 0 \right)$ &
$\left( \frac{2}{\sqrt{6}}\epsilon+\frac{1}{\sqrt{2}}\epsilon^2, -\sqrt{2}\epsilon -\frac{1}{\sqrt{6}}\epsilon^2, 0 \right)$ \\
$\epsilon_6$ & 
$\left( 0, \frac{1}{\sqrt{3}}\epsilon -\frac{2}{\sqrt{6}}\epsilon^2, -\frac{4}{\sqrt{6}}\epsilon-\frac{\sqrt{3}}{6} \epsilon^2 \right)$ &
$\left(0, \frac{1}{\sqrt{3}}\epsilon+\frac{1}{\sqrt{6}}\epsilon^2,\frac{2}{\sqrt{6}}\epsilon - \frac{\sqrt{3}}{6} \epsilon^2 \right)$ &
$\left( 0, \frac{1}{\sqrt{3}}\epsilon+\frac{1}{\sqrt{6}}\epsilon^2, \frac{2}{\sqrt{6}}\epsilon -\frac{\sqrt{3}}{6}\epsilon^2 \right)$ \\
$\epsilon_7$ &
$\left( \frac{1}{\sqrt{3}}\epsilon, 0, -\frac{\sqrt{3}}{6}\epsilon^2 \right)$ &
$\left( \frac{1}{\sqrt{3}}\epsilon+\frac{1}{\sqrt{2}}, 0, \sqrt{2}\epsilon -\frac{\sqrt{3}}{6}\epsilon^2 \right)$ &
$\left( \frac{1}{\sqrt{3}}\epsilon-\frac{1}{\sqrt{2}}\epsilon^2, 0, -\sqrt{2}\epsilon -\frac{\sqrt{3}}{6}\epsilon^2 \right)$ \\
\hline
\hline
\end{tabular}
\caption{\label{table:reciprocal_deformed_111} 
The reciprocal lattice vectors for the $(1,1,1)$-orientation.
The deformation gradient tensor and the group generators 
corresponding to the infinitesimal parameters $\{\epsilon_i\}$ 
are defined in Eqs.~\eqref{eq:DGT} and~\eqref{eq:SL3_generators}.}
\end{table*}

\section{Fourier transform of the magnetic dipole-dipole interaction}
\label{sec:FT_magnetic}
\subsection{Undeformed system}
In the case of the $\mathbf{k} \neq 0$ terms, we calculate the Fourier transform utilizing the plane wave expansion
\begin{align}
e^{i\mathbf{k}\cdot \mathbf{r}} = 4\pi \sum_{l=0}^\infty \sum_{m=-l}^l i^l j_l (kr) Y_l^m(\hat{k}) Y_l^{m*}(\hat{r}),
\end{align}
where $j_l$ and $Y_l^m$ are spherical Bessel functions and spherical harmonics, respectively, and the superscript asterisk $*$ denotes complex conjugate.
Since the dipole-dipole interaction energy [Eq.~\eqref{eq:u_m}] is proportional to $Y_2^0$ for $\mathbf{m} = m\hat{z}$, i.e., 
\begin{align}
u_m (\mathbf{r}) = \frac{\mu_0 m^2}{4\pi r^3} \left( -4 \sqrt{\frac{\pi}{5}} \right) Y_2^0 (\theta, \phi),
\end{align}
we obtain
\begin{align}
\tilde{u}_{\rm m}(\mathbf{k}) &= -\int_\sigma^\infty {\rm d}r \int_0^\pi {\rm d}\theta \int_0^{2\pi} {\rm d}\phi \,
	\frac{\mu_0 m^2}{\pi} \sqrt{\frac{\pi}{5}} \frac{\sin{\theta}}{r} Y_2^0 (\theta, \phi) e^{-i\mathbf{k}\cdot \mathbf{r}} 
	= 4 \mu_0 m^2 \sqrt{\frac{\pi}{5}} Y_2^0 (\theta_k , \phi_k ) \int_\sigma^\infty {\rm d}r\, \frac{j_2 (kr)}{r} \nonumber \\
	&= 4 \mu_0 m^2 \sqrt{\frac{\pi}{5}} (1- 3\cos^2 {\theta_k} ) \frac{k\sigma \cos {k\sigma} -\sin{k\sigma}}{k^3 \sigma^3}.
\end{align}

Meanwhile, the  $\mathbf{k}=0$ term is calculated as follows:
\begin{align}
\tilde{u}(\mathbf{k}=0)=&\lim_{R \to \infty} \frac{\mu_0 m^2}{4\pi} 2\pi \int_0^\pi {\rm d}\theta \int_\sigma^{{\gamma R}/{\sqrt{\cos^2{\theta} +\gamma^2 \sin^2 {\theta}}}} {\rm d} r
\,r^2 \sin{\theta} \frac{1-3\cos^2 {\theta}}{r^3} \nn \\
=&\lim_{R \to \infty} \frac{\mu_0 m^2}{2} \int_0^\pi {\rm d}\theta \, \sin{\theta}(1-3\cos^2{\theta}) 
	\ln{r}|_\sigma^{\gamma R / \sqrt{\cos^2{\theta} +\gamma^2 \sin^2 {\theta}}} \nn \\
=&-\frac{\mu_0 m^2}{2} \int_0^\pi {\rm d}\theta \,\sin{\theta}\, (1-3\cos^2{\theta})\, \ln{\sqrt{\cos^2 {\theta} +\gamma^2 \sin^2 {\theta}}} \nn \\
= &\begin{cases}
\displaystyle
-\frac{\mu_0 m^2}{2}
	\left[ 
		\frac{2}{3}+\frac{2}{\gamma^2 -1} 
		-\frac{ \gamma }{(\gamma^2 -1)^{3/2}}
		\left(
			\sinh^{-1} {\sqrt{\gamma^2 -1}} +\tanh^{-1} {\sqrt{\frac{\gamma^2 -1}{\gamma^2}}}
		\right)
	\right], & \gamma > 1. \\
\displaystyle
0, & \gamma = 1. \\
\displaystyle
- \frac{\mu_0 m^2}{2}
	\left[ 
		\frac{2}{3}+\frac{2}{\gamma^2 -1} 
		+\frac{ \gamma }{(1-\gamma^2)^{3/2}}
		\left(
			\sin^{-1} {\sqrt{1-\gamma^2}} +\tan^{-1} {\sqrt{\frac{1-\gamma^2}{\gamma^2}}}
		\right)
	\right], & \gamma < 1. \label{eq:u_k0_undeformed}
\end{cases}
\end{align}
where $\gamma = R_{\rm asp} R_{\rm sh}$.
Apparently, the $\mathbf{k}=0$ Fourier mode depends on the shape of systems, 
namely the aspect ratio $\gamma$.

\subsection{Deformed systems}
When $\mathbf{k} \neq 0$, the Fourier transformation is shape independent.
For the $\mathbf{k} =0$ mode, however, the Fourier transform of the deformed system is in general different 
from the undeformed one, due to the dependence on the sample shape. To calculate the correction, we first clarify how 
a deformation $\mathbf{F}$ modifies the integration via
\begin{align}
\int_{\Omega (r, \theta, \phi)} {\rm d}^3 r\, \frac{r^2 -3z^2}{r^5} 
	\to \int_{\Omega' (r, \theta, \phi)} {\rm d}^3 r\, \frac{r^2 -3z^2}{r^5},
\end{align}
where the prime indicates that the region of integration has been changed according to the deformation.
Then, we recover the original shape of the system by changing the variables via $\mathbf{r}' = \mathbf{F}\cdot \mathbf{r}$
where the center dot $\cdot$ denotes matrix multiplication, and subsequently, rewriting the integration as
\begin{align} \label{eq:ft_um}
	\int_0^\pi {\rm d}\theta \int_0^{2\pi} {\rm d} \phi \int_{\sigma' (\theta, \phi)}^{\gamma R_c /\sqrt{\cos^2 {\theta} +\gamma^2 \sin^2 {\theta}}} {\rm d} r\, \frac{r^2 \sin{\theta}\, (x'^2 +y'^2 - 2z'^2)}{r'^2} 
= \int_0^\pi {\rm d}\theta\, U(\theta),
\end{align}
while the boundaries of integration region stemming from the hard-core repulsion must be modified accordingly.
We note that the differential ${\rm d}^3 r = {\rm d}r\,r^2 \sin{\theta}$ remains unchanged because $|{\rm det}\, \mathbf{F}| = 1$. 
Also note that $\mathbf{F} \equiv {\rm d}\mathbf{r}'/{\rm d}\mathbf{r}$.
Then the above integration can be performed up to the second order of $\{\epsilon_i\}$  
with straightforward algebra, which has been performed using \emph{Mathematica}~\cite{mathematica}. Here, with 
\begin{align}
U_0 (\theta) = -\frac{\sin {\theta}  - 3\sin{3\theta} }{16} \ln{(\cos^2 {\theta} + \gamma^2 \sin^2 {\theta})},
\end{align}
we write the integrand in Eq.~\eqref{eq:ft_um} as
\begin{align}
U(\theta) = U_0 (\theta) +\Delta U (\theta, \{\epsilon_i\}),
\end{align}
where $\Delta U$ is the correction due to deformation.

First, for stretches (compressions) along the $x$-axis, combined with compressions (stretches) along the $y$-axis associated 
with $\boldsymbol{\lambda}_1$, $\Delta U$ reads
\begin{align} \label{eq:um_e1}
\Delta U(\theta, \epsilon_1) 
&=-\frac{\epsilon_1^2}{128}\sin^3 {\theta} [ 24+32\cos{2\theta}+72\cos{4\theta} +( 99+180\cos{2\theta} +105\cos{4\theta})\ln{(\cos^2 {\theta} +\gamma^2 \sin^2 {\theta})} ] \nn \\
&\equiv \Delta U_1 (\theta, \epsilon_1). 
\end{align} 
At the same time,
the correction stemming from the deformation associates with $\boldsymbol{\lambda}_2$ are already reflected in Eq.~\eqref{eq:u_k0_undeformed},
as we have calculated the values of $u_m (\mathbf{k}=0)$ for arbitrary aspect ratios. 
Because of the uniaxial symmetry of the magnetic dipolar interaction, 
the correction due to the shear deformations in the $xy$-plane takes the same form as Eq.~\eqref{eq:um_e1}, 
namely, $\Delta U(\theta, \epsilon_5) = \Delta U_1 (\theta, \epsilon_5)$. 
Next, the correction due to the shear deformations in the $yz$-plane is given as
\begin{align}
\Delta U(\theta, \epsilon_3)=& 
	\frac{\epsilon_3^2}{256}[
		36\sin{3\theta} + 28 \sin{5\theta} - 72\sin{7\theta} 
		+(15\sin{\theta} - 27\sin{3\theta} + 45\sin{5\theta}- 105\sin{7\theta}) 
			\ln{(\cos^2 {\theta} + \gamma^2 \sin^2 {\theta})}] \nn \\
\equiv & \Delta U_3 (\theta, \epsilon_3).
\end{align}
Due to the symmetry, we obtain $\Delta U(\theta, \epsilon_4) = \Delta U_3 (\theta, \epsilon_4)$ 
for the shear deformation in the $xz$-plane. 
Now we turn to the deformations involving rotations.
Again due to the uniaxial symmetry, 
the corrections due to the rotations in the $xz$- and $yz$-planes are identical with each other, reading
\begin{align}
\Delta U(\theta, \epsilon_6) =& \frac{3\epsilon_6^2}{32} (\sin{\theta} - 3\sin{3\theta}) \ln{(\cos^2 {\theta} + \gamma^2 \sin^2 {\theta})} \nn \\
\equiv & \Delta U_6 (\theta, \epsilon_6).
\end{align}
For the off-diagonal terms, 
the form of $\Delta U$ is simply given as the sum of two deformations,
except for the cases of $C_{36}$ and $C_{47}$,
in which the correction terms are given by
\begin{align}
\Delta U(\theta, \epsilon_3, \epsilon_6)=& 
	\Delta U_3 (\theta, \epsilon_3) + \Delta U_6 (\theta, \epsilon_6) \nn \\
&+\frac{\epsilon_3 \epsilon_6}{128} [ 
	28 \sin{\theta} + 18 \sin{3\theta} - 42 \sin{5\theta} 
	+ (6 \sin{\theta} + 45 \sin{3\theta} - 105\sin{5\theta})
		\ln{(\cos^2 {\theta} + \gamma^2 \sin^2 {\theta})}] \nn \\
\equiv &U_{36} (\theta, \epsilon_3, \epsilon_6),
\end{align}
and 
\begin{align}
\Delta U (\theta, \epsilon_4, \epsilon_7) = \Delta U_{36} (\theta, \epsilon_4, \epsilon_7).
\end{align}
The appearance of the additional correction terms of $\epsilon_3 \epsilon_6$ and $\epsilon_4 \epsilon_7$ 
is the direct consequence of the uniaxial symmetry underlying the magnetic dipole-dipole interaction. 
Such corrections correspond to the only nonzero mixing terms associated with the shear deformation and the rotation in any plane parallel to the anisotropy axis of uniaxial systems, namely the coefficient $D_2$ in Ref.~\onlinecite{Menzel2014}.
Moreover, equivalently to elastic constants as discussed in Appendix~\ref{sec:reciprocal}, the corrections to $\tilde{u}(\mathbf{k}=0)$ associated with mixing in the second order do not depend on nonlinear corrections in the deformation gradient $\mathbf{F}$. For the incompressibility constraint, they only appear at higher orders. 

\section{Fourier transform of the anisotropic pseudospring potential}
\label{sec:FT_pseudo}
When $\mathbf{k} = 0$, the Fourier transformation can be performed analytically.
In cylindrical coordinates, it reads
\begin{align}
\tilde{u}_{\rm pseudo} (\mathbf{k}=0)=&2\pi \int_{-R_{\rm asp} R_c}^{R_{\rm asp} R_c}{\rm d}z 
	\int_0^{\sqrt{R_c^2-z^2/ R_{\rm asp}^2}} {\rm d}\rho\,
	\rho \left[ \frac{1}{2}k_{\rm el} \left( \sqrt{\rho^2 +z^2}-a \right)^2 -u_0 \right] \nn \\
	=&2\pi \int_{-R_{\rm asp} R_c}^{R_{\rm asp} R_c}{\rm d}z \,
	\left[ \frac{1}{8} k_{\rm el}\rho^4 -\frac{1}{3}k_{\rm el}a (\rho^2 +z^2)^{3/2} 
	+\frac{1}{2}\rho^2 \left( \frac{1}{2}k_{\rm el} z^2 +\frac{1}{2}k_{\rm el}a^2 -u_0 \right) \right]_0^{\sqrt{R_c^2 -z^2/R_{\rm asp}^2}} \nn \\
	=&4\pi \int_0^{R_{\rm asp} R_c} {\rm d}z\, \left[ 
		\frac{1}{8}k_{\rm el} (R_c^2 -z^2/R_{\rm asp}^2 )^2 
		-\frac{1}{3}k_{\rm el}a (R_c^2-z^2/R_{\rm asp}^2 +z^2)^{3/2} +\frac{1}{3}k_{\rm el}a z^3
		\right. \nn \\
		&\hspace{3cm} \left. 
		+\frac{1}{2}\left( \frac{1}{2}k_{\rm el}z^2 +\frac{1}{2}k_{\rm el}a^2 -u_0 \right) (R_c^2 -z^2/R_{\rm asp}^2)
	\right],
\end{align}
which can be evaluated straightforwardly, except for
\begin{align}
\int_0^{R_{\rm asp} R_c}{\rm d}z\,  (R_c^2 -z^2/R_{\rm asp}^2+z^2)^{3/2}-z^3  
	=\frac{1}{8}R_{\rm asp}^2 R_c^4 (3+2R_{\rm asp}^2) 
	+\frac{3}{8}R_{\rm asp} R_c^4 \frac{\sinh^{-1}{\sqrt{-1+R_{\rm asp}^2}}}{\sqrt{-1+R_{\rm asp}^2}}.
\end{align}
Altogether, we obtain
\begin{align}
\tilde{u}_{\rm pseudo}(\mathbf{k}=0) =\begin{cases}  
2\pi \bigg[ 
	\frac{1}{15}R_{\rm asp} (2+R_{\rm asp}^2) k_{\rm el} R_c^5 
	-\frac{1}{4} \left( R_{\rm asp}^2 + \frac{R_{\rm asp} \sinh^{-1} {\sqrt{-1+R_{\rm asp}^2}}}{\sqrt{-1+R_{\rm asp}^2}}\right)
		k_{\rm el}a R_c^4 \nn \\
	\hspace{2cm}+\frac{2}{3}R_{\rm asp} \left( \frac{1}{2}k_{\rm el}a^2 -u_0 \right) R_c^3 \bigg] -C_0, &\hspace{1cm}R_{\rm asp} > 1, \vspace{0.3cm}  \nn \\ 
	4\pi \left[ \frac{1}{10}k_{\rm el} R_c^5 -\frac{1}{4}k_{\rm el} R_c^4 a
		+\frac{1}{3}R_c^3 \left( \frac{1}{2}k_{\rm el}a^2 -u_0 \right) \right] -C_0, &\hspace{1cm}R_{\rm asp} = 1,
\vspace{0.3cm} \nn \\
2\pi \bigg[
	\frac{1}{15}R_{\rm asp} (2+R_{\rm asp}^2) k_{\rm el} R_c^5 
	-\frac{1}{4} \left( R_{\rm asp}^2 + \frac{R_{\rm asp} \sin^{-1} {\sqrt{1-R_{\rm asp}^2}}}{\sqrt{1-R_{\rm asp}^2}}\right)
		k_{\rm el}a R_c^4 \nn \\
	\hspace{2cm} +\frac{2}{3}R_{\rm asp} \left( \frac{1}{2}k_{\rm el}a^2 -u_0 \right) R_c^3 \bigg] -C_0, &\hspace{1cm}R_{\rm asp} < 1,
	\end{cases}
\end{align}
where
\begin{align}
C_0\equiv	4\pi \left[ \frac{1}{10}k_{\rm el}\sigma^5 -\frac{1}{4}k_{\rm el} \sigma^4 a
		+\frac{1}{3}\sigma^3 \left( \frac{1}{2}k_{\rm el}a^2 -u_0 \right) \right].
\end{align}

When $\mathbf{k}\neq 0$, we have
\begin{align}
\tilde{u}_{\rm pseudo} (\mathbf{k}) = & 
	\int_{-R_{\rm asp} R_c}^{R_{\rm asp} R_c} {\rm d}z \int_0^{\sqrt{R_c^2 -z^2/R_{\rm asp}^2}} {\rm d}\rho \int_0^{2\pi} {\rm d}\phi \,
		\rho \left[ \frac{1}{2} k_{\rm el} \left(\sqrt{\rho^2 +z^2}-a \right)^2 -u_0 \right] e^{-i\mathbf{k} \cdot \mathbf{r}}
		\nonumber \\
		&-\int_0^\sigma {\rm d}r \int_0^\pi {\rm d}\theta \int_0^{2\pi} {\rm d} \phi\,
			r^2 \sin{\theta} \left[ \frac{1}{2}k_{\rm el} (r-a)^2 -u_0 \right] e^{-i \mathbf{k}\cdot \mathbf{r}} \nn \\
	=&4\pi \int_0^{R_{\rm asp} R_c} {\rm d}z \int_0^{\sqrt{R_c^2 -z^2/R_{\rm asp}^2}} {\rm d}\rho\,
		\rho \left[ \frac{1}{2} k_{\rm el} \left(\sqrt{\rho^2 +z^2}-a \right)^2 -u_0 \right]\, \cos{(k_z z)}\,
			J_0 \left( \rho \sqrt{k_x^2 +k_y^2} \right) \nn \\
		&-\frac{2\pi}{k^5} \left[ 
			4k_{\rm el}k + \left\{ k_{\rm el} (-4+6\sigma) k - k_{\rm el} (\sigma^3 -2\sigma^2 +\sigma) k^3 +2 u_0 \sigma k^3  \right\} \cos{k\sigma} \right. \nn \\ 
			&\hspace{1cm}\left. +\left\{ -6k_{\rm el}  +k_{\rm el} (1-4\sigma +3\sigma^2) k^2 -2 u_0 k^2 \right\} \sin{k\sigma} 
		\right].
\end{align}
We implemented the double integration in the third line using the \emph{Cubature} package~\cite{Johnson2017}.

We recall our assumption and observation underlying the mapping onto the pseudo-spring system
that the interaction does not depend much on the precise location of the boundaries of the Wigner-Seitz cells.
This results from the localization of the peaks in the density profiles mainly in the center of the cells~\cite{Goh2019}. 
Therefore, in contrast to the magnetic dipole-dipole interaction, 
we simply use the same value of $\tilde{u}_{\rm pseudo} (\mathbf{k})$ obtained for the undeformed geometry also for the deformed systems. 
\end{widetext}

\bibliography{3D_fgel}

\begin{thebibliography}{97}%
\makeatletter
\providecommand \@ifxundefined [1]{%
 \@ifx{#1\undefined}
}%
\providecommand \@ifnum [1]{%
 \ifnum #1\expandafter \@firstoftwo
 \else \expandafter \@secondoftwo
 \fi
}%
\providecommand \@ifx [1]{%
 \ifx #1\expandafter \@firstoftwo
 \else \expandafter \@secondoftwo
 \fi
}%
\providecommand \natexlab [1]{#1}%
\providecommand \enquote  [1]{``#1''}%
\providecommand \bibnamefont  [1]{#1}%
\providecommand \bibfnamefont [1]{#1}%
\providecommand \citenamefont [1]{#1}%
\providecommand \href@noop [0]{\@secondoftwo}%
\providecommand \href [0]{\begingroup \@sanitize@url \@href}%
\providecommand \@href[1]{\@@startlink{#1}\@@href}%
\providecommand \@@href[1]{\endgroup#1\@@endlink}%
\providecommand \@sanitize@url [0]{\catcode `\\12\catcode `\$12\catcode
  `\&12\catcode `\#12\catcode `\^12\catcode `\_12\catcode `\%12\relax}%
\providecommand \@@startlink[1]{}%
\providecommand \@@endlink[0]{}%
\providecommand \url  [0]{\begingroup\@sanitize@url \@url }%
\providecommand \@url [1]{\endgroup\@href {#1}{\urlprefix }}%
\providecommand \urlprefix  [0]{URL }%
\providecommand \Eprint [0]{\href }%
\providecommand \doibase [0]{http://dx.doi.org/}%
\providecommand \selectlanguage [0]{\@gobble}%
\providecommand \bibinfo  [0]{\@secondoftwo}%
\providecommand \bibfield  [0]{\@secondoftwo}%
\providecommand \translation [1]{[#1]}%
\providecommand \BibitemOpen [0]{}%
\providecommand \bibitemStop [0]{}%
\providecommand \bibitemNoStop [0]{.\EOS\space}%
\providecommand \EOS [0]{\spacefactor3000\relax}%
\providecommand \BibitemShut  [1]{\csname bibitem#1\endcsname}%
\let\auto@bib@innerbib\@empty
\bibitem [{\citenamefont {Filipcsei}\ \emph {et~al.}(2007)\citenamefont
  {Filipcsei}, \citenamefont {Csetneki}, \citenamefont {Szil{\'a}gyi},\ and\
  \citenamefont {Zr{\'i}nyi}}]{Filipcsei2007}%
  \BibitemOpen
  \bibfield  {author} {\bibinfo {author} {\bibfnamefont {G.}~\bibnamefont
  {Filipcsei}}, \bibinfo {author} {\bibfnamefont {I.}~\bibnamefont {Csetneki}},
  \bibinfo {author} {\bibfnamefont {A.}~\bibnamefont {Szil{\'a}gyi}}, \ and\
  \bibinfo {author} {\bibfnamefont {M.}~\bibnamefont {Zr{\'i}nyi}},\ }\href
  {\doibase 10.1007/12_2006_104} {\bibfield  {journal} {\bibinfo  {journal}
  {Adv. Polym. Sci.}\ }\textbf {\bibinfo {volume} {206}},\ \bibinfo {pages}
  {137} (\bibinfo {year} {2007})}\BibitemShut {NoStop}%
\bibitem [{\citenamefont {Ilg}(2013)}]{Ilg2013}%
  \BibitemOpen
  \bibfield  {author} {\bibinfo {author} {\bibfnamefont {P.}~\bibnamefont
  {Ilg}},\ }\href {\doibase 10.1039/C3SM27809C} {\bibfield  {journal} {\bibinfo
   {journal} {Soft Matter}\ }\textbf {\bibinfo {volume} {9}},\ \bibinfo {pages}
  {3465} (\bibinfo {year} {2013})}\BibitemShut {NoStop}%
\bibitem [{\citenamefont {Menzel}(2015)}]{Menzel2015}%
  \BibitemOpen
  \bibfield  {author} {\bibinfo {author} {\bibfnamefont {A.~M.}\ \bibnamefont
  {Menzel}},\ }\href {\doibase 10.1016/j.physrep.2014.10.001} {\bibfield
  {journal} {\bibinfo  {journal} {Phys. Rep.}\ }\textbf {\bibinfo {volume}
  {554}},\ \bibinfo {pages} {1 } (\bibinfo {year} {2015})}\BibitemShut
  {NoStop}%
\bibitem [{\citenamefont {Odenbach}(2016)}]{Odenbach2016}%
  \BibitemOpen
  \bibfield  {author} {\bibinfo {author} {\bibfnamefont {S.}~\bibnamefont
  {Odenbach}},\ }\href {\doibase 10.1007/s00419-015-1092-6} {\bibfield
  {journal} {\bibinfo  {journal} {Arch. Appl. Mech.}\ }\textbf {\bibinfo
  {volume} {86}},\ \bibinfo {pages} {269} (\bibinfo {year} {2016})}\BibitemShut
  {NoStop}%
\bibitem [{\citenamefont {Frickel}, \citenamefont {Messing},\ and\
  \citenamefont {Schmidt}(2011)}]{Frickel2011}%
  \BibitemOpen
  \bibfield  {author} {\bibinfo {author} {\bibfnamefont {N.}~\bibnamefont
  {Frickel}}, \bibinfo {author} {\bibfnamefont {R.}~\bibnamefont {Messing}}, \
  and\ \bibinfo {author} {\bibfnamefont {A.~M.}\ \bibnamefont {Schmidt}},\
  }\href {\doibase 10.1039/C0JM03816D} {\bibfield  {journal} {\bibinfo
  {journal} {J. Mater. Chem.}\ }\textbf {\bibinfo {volume} {21}},\ \bibinfo
  {pages} {8466} (\bibinfo {year} {2011})}\BibitemShut {NoStop}%
\bibitem [{\citenamefont {Messing}\ \emph {et~al.}(2011)\citenamefont
  {Messing}, \citenamefont {Frickel}, \citenamefont {Belkoura}, \citenamefont
  {Strey}, \citenamefont {Rahn}, \citenamefont {Odenbach},\ and\ \citenamefont
  {Schmidt}}]{Messing2011}%
  \BibitemOpen
  \bibfield  {author} {\bibinfo {author} {\bibfnamefont {R.}~\bibnamefont
  {Messing}}, \bibinfo {author} {\bibfnamefont {N.}~\bibnamefont {Frickel}},
  \bibinfo {author} {\bibfnamefont {L.}~\bibnamefont {Belkoura}}, \bibinfo
  {author} {\bibfnamefont {R.}~\bibnamefont {Strey}}, \bibinfo {author}
  {\bibfnamefont {H.}~\bibnamefont {Rahn}}, \bibinfo {author} {\bibfnamefont
  {S.}~\bibnamefont {Odenbach}}, \ and\ \bibinfo {author} {\bibfnamefont
  {A.~M.}\ \bibnamefont {Schmidt}},\ }\href {\doibase 10.1021/ma102708b}
  {\bibfield  {journal} {\bibinfo  {journal} {Macromolecules}\ }\textbf
  {\bibinfo {volume} {44}},\ \bibinfo {pages} {2990} (\bibinfo {year}
  {2011})}\BibitemShut {NoStop}%
\bibitem [{\citenamefont {Gundermann}\ and\ \citenamefont
  {Odenbach}(2014)}]{Gundermann2014}%
  \BibitemOpen
  \bibfield  {author} {\bibinfo {author} {\bibfnamefont {T.}~\bibnamefont
  {Gundermann}}\ and\ \bibinfo {author} {\bibfnamefont {S.}~\bibnamefont
  {Odenbach}},\ }\href {\doibase 10.1088/0964-1726/23/10/105013} {\bibfield
  {journal} {\bibinfo  {journal} {Smart Mater. Struct.}\ }\textbf {\bibinfo
  {volume} {23}},\ \bibinfo {pages} {105013} (\bibinfo {year}
  {2014})}\BibitemShut {NoStop}%
\bibitem [{\citenamefont {Landers}\ \emph {et~al.}(2015)\citenamefont
  {Landers}, \citenamefont {Roeder}, \citenamefont {Salamon}, \citenamefont
  {Schmidt},\ and\ \citenamefont {Wende}}]{Landers2015}%
  \BibitemOpen
  \bibfield  {author} {\bibinfo {author} {\bibfnamefont {J.}~\bibnamefont
  {Landers}}, \bibinfo {author} {\bibfnamefont {L.}~\bibnamefont {Roeder}},
  \bibinfo {author} {\bibfnamefont {S.}~\bibnamefont {Salamon}}, \bibinfo
  {author} {\bibfnamefont {A.~M.}\ \bibnamefont {Schmidt}}, \ and\ \bibinfo
  {author} {\bibfnamefont {H.}~\bibnamefont {Wende}},\ }\href {\doibase
  10.1021/acs.jpcc.5b03697} {\bibfield  {journal} {\bibinfo  {journal} {J.
  Phys. Chem. C}\ }\textbf {\bibinfo {volume} {119}},\ \bibinfo {pages} {20642}
  (\bibinfo {year} {2015})}\BibitemShut {NoStop}%
\bibitem [{\citenamefont {Roeder}\ \emph {et~al.}(2015)\citenamefont {Roeder},
  \citenamefont {Bender}, \citenamefont {Kundt}, \citenamefont {Tsch{\"o}pe},\
  and\ \citenamefont {Schmidt}}]{Roeder2015}%
  \BibitemOpen
  \bibfield  {author} {\bibinfo {author} {\bibfnamefont {L.}~\bibnamefont
  {Roeder}}, \bibinfo {author} {\bibfnamefont {P.}~\bibnamefont {Bender}},
  \bibinfo {author} {\bibfnamefont {M.}~\bibnamefont {Kundt}}, \bibinfo
  {author} {\bibfnamefont {A.}~\bibnamefont {Tsch{\"o}pe}}, \ and\ \bibinfo
  {author} {\bibfnamefont {A.~M.}\ \bibnamefont {Schmidt}},\ }\href {\doibase
  10.1039/C4CP04493B} {\bibfield  {journal} {\bibinfo  {journal} {Phys. Chem.
  Chem. Phys.}\ }\textbf {\bibinfo {volume} {17}},\ \bibinfo {pages} {1290}
  (\bibinfo {year} {2015})}\BibitemShut {NoStop}%
\bibitem [{\citenamefont {Ginder}\ \emph {et~al.}(2002)\citenamefont {Ginder},
  \citenamefont {Clark}, \citenamefont {Schlotter},\ and\ \citenamefont
  {Nichols}}]{Ginder2002}%
  \BibitemOpen
  \bibfield  {author} {\bibinfo {author} {\bibfnamefont {J.~M.}\ \bibnamefont
  {Ginder}}, \bibinfo {author} {\bibfnamefont {S.~M.}\ \bibnamefont {Clark}},
  \bibinfo {author} {\bibfnamefont {W.~F.}\ \bibnamefont {Schlotter}}, \ and\
  \bibinfo {author} {\bibfnamefont {M.~E.}\ \bibnamefont {Nichols}},\ }\href
  {\doibase 10.1142/S021797920201244X} {\bibfield  {journal} {\bibinfo
  {journal} {Int. J. Mod. Phys. B}\ }\textbf {\bibinfo {volume} {16}},\
  \bibinfo {pages} {2412} (\bibinfo {year} {2002})}\BibitemShut {NoStop}%
\bibitem [{\citenamefont {Zhou}\ and\ \citenamefont {Jiang}(2004)}]{Zhou2004}%
  \BibitemOpen
  \bibfield  {author} {\bibinfo {author} {\bibfnamefont {G.~Y.}\ \bibnamefont
  {Zhou}}\ and\ \bibinfo {author} {\bibfnamefont {Z.~Y.}\ \bibnamefont
  {Jiang}},\ }\href {\doibase 10.1088/0964-1726/13/2/009} {\bibfield  {journal}
  {\bibinfo  {journal} {Smart Mater. Struct.}\ }\textbf {\bibinfo {volume}
  {13}},\ \bibinfo {pages} {309} (\bibinfo {year} {2004})}\BibitemShut
  {NoStop}%
\bibitem [{\citenamefont {Gollwitzer}\ \emph {et~al.}(2008)\citenamefont
  {Gollwitzer}, \citenamefont {Turanov}, \citenamefont {Krekhova},
  \citenamefont {Lattermann}, \citenamefont {Rehberg},\ and\ \citenamefont
  {Richter}}]{Gollwitzer2008}%
  \BibitemOpen
  \bibfield  {author} {\bibinfo {author} {\bibfnamefont {C.}~\bibnamefont
  {Gollwitzer}}, \bibinfo {author} {\bibfnamefont {A.}~\bibnamefont {Turanov}},
  \bibinfo {author} {\bibfnamefont {M.}~\bibnamefont {Krekhova}}, \bibinfo
  {author} {\bibfnamefont {G.}~\bibnamefont {Lattermann}}, \bibinfo {author}
  {\bibfnamefont {I.}~\bibnamefont {Rehberg}}, \ and\ \bibinfo {author}
  {\bibfnamefont {R.}~\bibnamefont {Richter}},\ }\href {\doibase
  10.1063/1.2905212} {\bibfield  {journal} {\bibinfo  {journal} {J. Chem.
  Phys.}\ }\textbf {\bibinfo {volume} {128}},\ \bibinfo {pages} {164709}
  (\bibinfo {year} {2008})}\BibitemShut {NoStop}%
\bibitem [{\citenamefont {Borin}, \citenamefont {Stepanov},\ and\ \citenamefont
  {Odenbach}(2013)}]{Borin2013}%
  \BibitemOpen
  \bibfield  {author} {\bibinfo {author} {\bibfnamefont {D.~Y.}\ \bibnamefont
  {Borin}}, \bibinfo {author} {\bibfnamefont {G.~V.}\ \bibnamefont {Stepanov}},
  \ and\ \bibinfo {author} {\bibfnamefont {S.}~\bibnamefont {Odenbach}},\
  }\href {\doibase 10.1088/1742-6596/412/1/012040} {\bibfield  {journal}
  {\bibinfo  {journal} {J. Phys.: Conf. Ser.}\ }\textbf {\bibinfo {volume}
  {412}},\ \bibinfo {pages} {012040} (\bibinfo {year} {2013})}\BibitemShut
  {NoStop}%
\bibitem [{\citenamefont {Kramarenko}\ \emph {et~al.}(2015)\citenamefont
  {Kramarenko}, \citenamefont {Chertovich}, \citenamefont {Stepanov},
  \citenamefont {Semisalova}, \citenamefont {Makarova}, \citenamefont {Perov},\
  and\ \citenamefont {Khokhlov}}]{Kramarenko2015}%
  \BibitemOpen
  \bibfield  {author} {\bibinfo {author} {\bibfnamefont {E.~Y.}\ \bibnamefont
  {Kramarenko}}, \bibinfo {author} {\bibfnamefont {A.~V.}\ \bibnamefont
  {Chertovich}}, \bibinfo {author} {\bibfnamefont {G.~V.}\ \bibnamefont
  {Stepanov}}, \bibinfo {author} {\bibfnamefont {A.~S.}\ \bibnamefont
  {Semisalova}}, \bibinfo {author} {\bibfnamefont {L.~A.}\ \bibnamefont
  {Makarova}}, \bibinfo {author} {\bibfnamefont {N.~S.}\ \bibnamefont {Perov}},
  \ and\ \bibinfo {author} {\bibfnamefont {A.~R.}\ \bibnamefont {Khokhlov}},\
  }\href {\doibase 10.1088/0964-1726/24/3/035002} {\bibfield  {journal}
  {\bibinfo  {journal} {Smart Mater. Struct.}\ }\textbf {\bibinfo {volume}
  {24}},\ \bibinfo {pages} {035002} (\bibinfo {year} {2015})}\BibitemShut
  {NoStop}%
\bibitem [{\citenamefont {Safronov}\ and\ \citenamefont
  {Mikhnevich}(2019)}]{Safronov2019}%
  \BibitemOpen
  \bibfield  {author} {\bibinfo {author} {\bibfnamefont {A.~P.}\ \bibnamefont
  {Safronov}}\ and\ \bibinfo {author} {\bibfnamefont {E.~A.}\ \bibnamefont
  {Mikhnevich}},\ }\href {\doibase 10.1088/1742-6596/1389/1/012057} {\bibfield
  {journal} {\bibinfo  {journal} {J. Phys.: Conf. Ser.}\ }\textbf {\bibinfo
  {volume} {1389}},\ \bibinfo {pages} {012057} (\bibinfo {year}
  {2019})}\BibitemShut {NoStop}%
\bibitem [{\citenamefont {Borin}, \citenamefont {Odenbach},\ and\ \citenamefont
  {Stepanov}(2019)}]{Borin2019}%
  \BibitemOpen
  \bibfield  {author} {\bibinfo {author} {\bibfnamefont {D.}~\bibnamefont
  {Borin}}, \bibinfo {author} {\bibfnamefont {S.}~\bibnamefont {Odenbach}}, \
  and\ \bibinfo {author} {\bibfnamefont {G.}~\bibnamefont {Stepanov}},\ }\href
  {\doibase 10.1016/j.jmmm.2017.12.072} {\bibfield  {journal} {\bibinfo
  {journal} {J. Magn. Magn. Mater.}\ }\textbf {\bibinfo {volume} {470}},\
  \bibinfo {pages} {85 } (\bibinfo {year} {2019})}\BibitemShut {NoStop}%
\bibitem [{\citenamefont {Saveliev}\ \emph {et~al.}(2020)\citenamefont
  {Saveliev}, \citenamefont {Belyaeva}, \citenamefont {Chashin}, \citenamefont
  {Fetisov}, \citenamefont {Romeis}, \citenamefont {Kettl}, \citenamefont
  {Kramarenko}, \citenamefont {Saphiannikova}, \citenamefont {Stepanov},\ and\
  \citenamefont {Shamonin}}]{Saveliev2020}%
  \BibitemOpen
  \bibfield  {author} {\bibinfo {author} {\bibfnamefont {D.}~\bibnamefont
  {Saveliev}}, \bibinfo {author} {\bibfnamefont {I.}~\bibnamefont {Belyaeva}},
  \bibinfo {author} {\bibfnamefont {D.}~\bibnamefont {Chashin}}, \bibinfo
  {author} {\bibfnamefont {L.}~\bibnamefont {Fetisov}}, \bibinfo {author}
  {\bibfnamefont {D.}~\bibnamefont {Romeis}}, \bibinfo {author} {\bibfnamefont
  {W.}~\bibnamefont {Kettl}}, \bibinfo {author} {\bibfnamefont
  {E.}~\bibnamefont {Kramarenko}}, \bibinfo {author} {\bibfnamefont
  {M.}~\bibnamefont {Saphiannikova}}, \bibinfo {author} {\bibfnamefont
  {G.}~\bibnamefont {Stepanov}}, \ and\ \bibinfo {author} {\bibfnamefont
  {M.}~\bibnamefont {Shamonin}},\ }\href {\doibase 10.3390/ma13153297}
  {\bibfield  {journal} {\bibinfo  {journal} {Mater.}\ }\textbf {\bibinfo
  {volume} {13}},\ \bibinfo {pages} {3297} (\bibinfo {year}
  {2020})}\BibitemShut {NoStop}%
\bibitem [{\citenamefont {Danas}, \citenamefont {Kankanala},\ and\
  \citenamefont {Triantafyllidis}(2012)}]{Danas2012}%
  \BibitemOpen
  \bibfield  {author} {\bibinfo {author} {\bibfnamefont {K.}~\bibnamefont
  {Danas}}, \bibinfo {author} {\bibfnamefont {S.}~\bibnamefont {Kankanala}}, \
  and\ \bibinfo {author} {\bibfnamefont {N.}~\bibnamefont {Triantafyllidis}},\
  }\href {\doibase 10.1016/j.jmps.2011.09.006} {\bibfield  {journal} {\bibinfo
  {journal} {J. Mech. Phys. Solids}\ }\textbf {\bibinfo {volume} {60}},\
  \bibinfo {pages} {120 } (\bibinfo {year} {2012})}\BibitemShut {NoStop}%
\bibitem [{\citenamefont {Gundermann}\ \emph {et~al.}(2017)\citenamefont
  {Gundermann}, \citenamefont {Cremer}, \citenamefont {L\"owen}, \citenamefont
  {Menzel},\ and\ \citenamefont {Odenbach}}]{Gundermann2017}%
  \BibitemOpen
  \bibfield  {author} {\bibinfo {author} {\bibfnamefont {T.}~\bibnamefont
  {Gundermann}}, \bibinfo {author} {\bibfnamefont {P.}~\bibnamefont {Cremer}},
  \bibinfo {author} {\bibfnamefont {H.}~\bibnamefont {L\"owen}}, \bibinfo
  {author} {\bibfnamefont {A.~M.}\ \bibnamefont {Menzel}}, \ and\ \bibinfo
  {author} {\bibfnamefont {S.}~\bibnamefont {Odenbach}},\ }\href {\doibase
  10.1088/1361-665X/aa5f96} {\bibfield  {journal} {\bibinfo  {journal} {Smart
  Mater. Struct.}\ }\textbf {\bibinfo {volume} {26}},\ \bibinfo {pages}
  {045012} (\bibinfo {year} {2017})}\BibitemShut {NoStop}%
\bibitem [{\citenamefont {Pessot}\ \emph {et~al.}(2018)\citenamefont {Pessot},
  \citenamefont {Sch{\"u}mann}, \citenamefont {Gundermann}, \citenamefont
  {Odenbach}, \citenamefont {L{\"o}wen},\ and\ \citenamefont
  {Menzel}}]{Pessot2018}%
  \BibitemOpen
  \bibfield  {author} {\bibinfo {author} {\bibfnamefont {G.}~\bibnamefont
  {Pessot}}, \bibinfo {author} {\bibfnamefont {M.}~\bibnamefont
  {Sch{\"u}mann}}, \bibinfo {author} {\bibfnamefont {T.}~\bibnamefont
  {Gundermann}}, \bibinfo {author} {\bibfnamefont {S.}~\bibnamefont
  {Odenbach}}, \bibinfo {author} {\bibfnamefont {H.}~\bibnamefont {L{\"o}wen}},
  \ and\ \bibinfo {author} {\bibfnamefont {A.~M.}\ \bibnamefont {Menzel}},\
  }\href {\doibase 10.1088/1361-648X/aaaeaa} {\bibfield  {journal} {\bibinfo
  {journal} {J. Phys.: Condens. Matter}\ }\textbf {\bibinfo {volume} {30}},\
  \bibinfo {pages} {125101} (\bibinfo {year} {2018})}\BibitemShut {NoStop}%
\bibitem [{\citenamefont {Puljiz}\ \emph {et~al.}(2018)\citenamefont {Puljiz},
  \citenamefont {Huang}, \citenamefont {Kalina}, \citenamefont {Nowak},
  \citenamefont {Odenbach}, \citenamefont {K{\"a}stner}, \citenamefont
  {Auernhammer},\ and\ \citenamefont {Menzel}}]{Puljiz2018}%
  \BibitemOpen
  \bibfield  {author} {\bibinfo {author} {\bibfnamefont {M.}~\bibnamefont
  {Puljiz}}, \bibinfo {author} {\bibfnamefont {S.}~\bibnamefont {Huang}},
  \bibinfo {author} {\bibfnamefont {K.~A.}\ \bibnamefont {Kalina}}, \bibinfo
  {author} {\bibfnamefont {J.}~\bibnamefont {Nowak}}, \bibinfo {author}
  {\bibfnamefont {S.}~\bibnamefont {Odenbach}}, \bibinfo {author}
  {\bibfnamefont {M.}~\bibnamefont {K{\"a}stner}}, \bibinfo {author}
  {\bibfnamefont {G.~K.}\ \bibnamefont {Auernhammer}}, \ and\ \bibinfo {author}
  {\bibfnamefont {A.~M.}\ \bibnamefont {Menzel}},\ }\href {\doibase
  10.1039/C8SM01051J} {\bibfield  {journal} {\bibinfo  {journal} {Soft Matter}\
  }\textbf {\bibinfo {volume} {14}},\ \bibinfo {pages} {6809} (\bibinfo {year}
  {2018})}\BibitemShut {NoStop}%
\bibitem [{\citenamefont {Sturm}\ \emph {et~al.}(2019)\citenamefont {Sturm},
  \citenamefont {Siglreitmeier}, \citenamefont {Wolf}, \citenamefont {Vogel},
  \citenamefont {Gratz}, \citenamefont {Faivre}, \citenamefont {Lubk},
  \citenamefont {B{\"u}chner}, \citenamefont {Sturm},\ and\ \citenamefont
  {C{\"o}lfen}}]{Sturm2019}%
  \BibitemOpen
  \bibfield  {author} {\bibinfo {author} {\bibfnamefont {S.}~\bibnamefont
  {Sturm}}, \bibinfo {author} {\bibfnamefont {M.}~\bibnamefont
  {Siglreitmeier}}, \bibinfo {author} {\bibfnamefont {D.}~\bibnamefont {Wolf}},
  \bibinfo {author} {\bibfnamefont {K.}~\bibnamefont {Vogel}}, \bibinfo
  {author} {\bibfnamefont {M.}~\bibnamefont {Gratz}}, \bibinfo {author}
  {\bibfnamefont {D.}~\bibnamefont {Faivre}}, \bibinfo {author} {\bibfnamefont
  {A.}~\bibnamefont {Lubk}}, \bibinfo {author} {\bibfnamefont {B.}~\bibnamefont
  {B{\"u}chner}}, \bibinfo {author} {\bibfnamefont {E.~V.}\ \bibnamefont
  {Sturm}}, \ and\ \bibinfo {author} {\bibfnamefont {H.}~\bibnamefont
  {C{\"o}lfen}},\ }\href {\doibase 10.1002/adfm.201905996} {\bibfield
  {journal} {\bibinfo  {journal} {Adv. Funct. Mater.}\ }\textbf {\bibinfo
  {volume} {29}},\ \bibinfo {pages} {1905996} (\bibinfo {year}
  {2019})}\BibitemShut {NoStop}%
\bibitem [{\citenamefont {Sch{\"u}mann}\ \emph {et~al.}(2020)\citenamefont
  {Sch{\"u}mann}, \citenamefont {Morich}, \citenamefont {G{\"u}nther},\ and\
  \citenamefont {Odenbach}}]{Schumann2020}%
  \BibitemOpen
  \bibfield  {author} {\bibinfo {author} {\bibfnamefont {M.}~\bibnamefont
  {Sch{\"u}mann}}, \bibinfo {author} {\bibfnamefont {J.}~\bibnamefont
  {Morich}}, \bibinfo {author} {\bibfnamefont {S.}~\bibnamefont {G{\"u}nther}},
  \ and\ \bibinfo {author} {\bibfnamefont {S.}~\bibnamefont {Odenbach}},\
  }\href {\doibase 10.1016/j.jmmm.2020.166537} {\bibfield  {journal} {\bibinfo
  {journal} {J. Magn. Magn. Mater.}\ }\textbf {\bibinfo {volume} {502}},\
  \bibinfo {pages} {166537} (\bibinfo {year} {2020})}\BibitemShut {NoStop}%
\bibitem [{\citenamefont {Cremer}\ \emph {et~al.}(2017)\citenamefont {Cremer},
  \citenamefont {Heinen}, \citenamefont {Menzel},\ and\ \citenamefont
  {L{\"o}wen}}]{Cremer2017}%
  \BibitemOpen
  \bibfield  {author} {\bibinfo {author} {\bibfnamefont {P.}~\bibnamefont
  {Cremer}}, \bibinfo {author} {\bibfnamefont {M.}~\bibnamefont {Heinen}},
  \bibinfo {author} {\bibfnamefont {A.~M.}\ \bibnamefont {Menzel}}, \ and\
  \bibinfo {author} {\bibfnamefont {H.}~\bibnamefont {L{\"o}wen}},\ }\href
  {\doibase 10.1088/1361-648X/aa73bd} {\bibfield  {journal} {\bibinfo
  {journal} {J. Phys.: Condens. Matter}\ }\textbf {\bibinfo {volume} {29}},\
  \bibinfo {pages} {275102} (\bibinfo {year} {2017})}\BibitemShut {NoStop}%
\bibitem [{\citenamefont {Jarkova}\ \emph {et~al.}(2003)\citenamefont
  {Jarkova}, \citenamefont {Pleiner}, \citenamefont {M\"uller},\ and\
  \citenamefont {Brand}}]{Jarkova2003}%
  \BibitemOpen
  \bibfield  {author} {\bibinfo {author} {\bibfnamefont {E.}~\bibnamefont
  {Jarkova}}, \bibinfo {author} {\bibfnamefont {H.}~\bibnamefont {Pleiner}},
  \bibinfo {author} {\bibfnamefont {H.-W.}\ \bibnamefont {M\"uller}}, \ and\
  \bibinfo {author} {\bibfnamefont {H.~R.}\ \bibnamefont {Brand}},\ }\href
  {\doibase 10.1103/PhysRevE.68.041706} {\bibfield  {journal} {\bibinfo
  {journal} {Phys. Rev. E}\ }\textbf {\bibinfo {volume} {68}},\ \bibinfo
  {pages} {041706} (\bibinfo {year} {2003})}\BibitemShut {NoStop}%
\bibitem [{\citenamefont {Bohlius}, \citenamefont {Brand},\ and\ \citenamefont
  {Pleiner}(2004)}]{Bohlius2004}%
  \BibitemOpen
  \bibfield  {author} {\bibinfo {author} {\bibfnamefont {S.}~\bibnamefont
  {Bohlius}}, \bibinfo {author} {\bibfnamefont {H.~R.}\ \bibnamefont {Brand}},
  \ and\ \bibinfo {author} {\bibfnamefont {H.}~\bibnamefont {Pleiner}},\ }\href
  {\doibase 10.1103/PhysRevE.70.061411} {\bibfield  {journal} {\bibinfo
  {journal} {Phys. Rev. E}\ }\textbf {\bibinfo {volume} {70}},\ \bibinfo
  {pages} {061411} (\bibinfo {year} {2004})}\BibitemShut {NoStop}%
\bibitem [{\citenamefont {Potisk}, \citenamefont {Pleiner},\ and\ \citenamefont
  {Brand}(2019)}]{Potisk2019}%
  \BibitemOpen
  \bibfield  {author} {\bibinfo {author} {\bibfnamefont {T.}~\bibnamefont
  {Potisk}}, \bibinfo {author} {\bibfnamefont {H.}~\bibnamefont {Pleiner}}, \
  and\ \bibinfo {author} {\bibfnamefont {H.~R.}\ \bibnamefont {Brand}},\ }\href
  {\doibase 10.1140/epje/i2019-11798-6} {\bibfield  {journal} {\bibinfo
  {journal} {Euro. Phys. J. E}\ }\textbf {\bibinfo {volume} {42}},\ \bibinfo
  {pages} {35} (\bibinfo {year} {2019})}\BibitemShut {NoStop}%
\bibitem [{\citenamefont {Raikher}\ and\ \citenamefont
  {Stolbov}(2005)}]{Raikher2005}%
  \BibitemOpen
  \bibfield  {author} {\bibinfo {author} {\bibfnamefont {Y.}~\bibnamefont
  {Raikher}}\ and\ \bibinfo {author} {\bibfnamefont {O.}~\bibnamefont
  {Stolbov}},\ }\href {\doibase 10.1016/j.jmmm.2004.11.018} {\bibfield
  {journal} {\bibinfo  {journal} {J. Magn. Magn. Mater.}\ }\textbf {\bibinfo
  {volume} {289}},\ \bibinfo {pages} {62 } (\bibinfo {year}
  {2005})}\BibitemShut {NoStop}%
\bibitem [{\citenamefont {Raikher}\ and\ \citenamefont
  {Stolbov}(2008)}]{Raikher2008}%
  \BibitemOpen
  \bibfield  {author} {\bibinfo {author} {\bibfnamefont {Y.~L.}\ \bibnamefont
  {Raikher}}\ and\ \bibinfo {author} {\bibfnamefont {O.~V.}\ \bibnamefont
  {Stolbov}},\ }\href {\doibase 10.1088/0953-8984/20/20/204126} {\bibfield
  {journal} {\bibinfo  {journal} {J. Phys.: Condens. Matter}\ }\textbf
  {\bibinfo {volume} {20}},\ \bibinfo {pages} {204126} (\bibinfo {year}
  {2008})}\BibitemShut {NoStop}%
\bibitem [{\citenamefont {Stolbov}\ and\ \citenamefont
  {Raikher}(2019)}]{Stolbov2019}%
  \BibitemOpen
  \bibfield  {author} {\bibinfo {author} {\bibfnamefont {O.~V.}\ \bibnamefont
  {Stolbov}}\ and\ \bibinfo {author} {\bibfnamefont {Y.~L.}\ \bibnamefont
  {Raikher}},\ }\href {\doibase 10.1007/s00419-018-1452-0} {\bibfield
  {journal} {\bibinfo  {journal} {Arch. Appl. Mech.}\ }\textbf {\bibinfo
  {volume} {89}},\ \bibinfo {pages} {63} (\bibinfo {year} {2019})}\BibitemShut
  {NoStop}%
\bibitem [{\citenamefont {Weeber}, \citenamefont {Kantorovich},\ and\
  \citenamefont {Holm}(2012)}]{Weeber2012}%
  \BibitemOpen
  \bibfield  {author} {\bibinfo {author} {\bibfnamefont {R.}~\bibnamefont
  {Weeber}}, \bibinfo {author} {\bibfnamefont {S.~S.}\ \bibnamefont
  {Kantorovich}}, \ and\ \bibinfo {author} {\bibfnamefont {C.}~\bibnamefont
  {Holm}},\ }\href {\doibase 10.1039/C2SM26097B} {\bibfield  {journal}
  {\bibinfo  {journal} {Soft Matter}\ }\textbf {\bibinfo {volume} {8}},\
  \bibinfo {pages} {9923} (\bibinfo {year} {2012})}\BibitemShut {NoStop}%
\bibitem [{\citenamefont {Weeber}, \citenamefont {S.},\ and\ \citenamefont
  {Holm}(2015)}]{Weeber2015jcp}%
  \BibitemOpen
  \bibfield  {author} {\bibinfo {author} {\bibfnamefont {R.}~\bibnamefont
  {Weeber}}, \bibinfo {author} {\bibfnamefont {S.~K.}\ \bibnamefont {S.}}, \
  and\ \bibinfo {author} {\bibfnamefont {C.}~\bibnamefont {Holm}},\ }\href
  {\doibase 10.1063/1.4932371} {\bibfield  {journal} {\bibinfo  {journal} {J.
  Chem. Phys.}\ }\textbf {\bibinfo {volume} {143}},\ \bibinfo {pages} {154901}
  (\bibinfo {year} {2015})}\BibitemShut {NoStop}%
\bibitem [{\citenamefont {Minina}\ \emph {et~al.}(2018)\citenamefont {Minina},
  \citenamefont {S{\'a}nchez}, \citenamefont {Likos},\ and\ \citenamefont
  {Kantorovich}}]{Minina2018}%
  \BibitemOpen
  \bibfield  {author} {\bibinfo {author} {\bibfnamefont {E.~S.}\ \bibnamefont
  {Minina}}, \bibinfo {author} {\bibfnamefont {P.~A.}\ \bibnamefont
  {S{\'a}nchez}}, \bibinfo {author} {\bibfnamefont {C.~N.}\ \bibnamefont
  {Likos}}, \ and\ \bibinfo {author} {\bibfnamefont {S.~S.}\ \bibnamefont
  {Kantorovich}},\ }\href {\doibase 10.1016/j.jmmm.2017.10.107} {\bibfield
  {journal} {\bibinfo  {journal} {J. Magn. Magn. Mater.}\ }\textbf {\bibinfo
  {volume} {459}},\ \bibinfo {pages} {226 } (\bibinfo {year}
  {2018})}\BibitemShut {NoStop}%
\bibitem [{\citenamefont {Stolbov}, \citenamefont {Raikher},\ and\
  \citenamefont {Balasoiu}(2011)}]{Stolbov2011}%
  \BibitemOpen
  \bibfield  {author} {\bibinfo {author} {\bibfnamefont {O.~V.}\ \bibnamefont
  {Stolbov}}, \bibinfo {author} {\bibfnamefont {Y.~L.}\ \bibnamefont
  {Raikher}}, \ and\ \bibinfo {author} {\bibfnamefont {M.}~\bibnamefont
  {Balasoiu}},\ }\href {\doibase 10.1039/C1SM05714F} {\bibfield  {journal}
  {\bibinfo  {journal} {Soft Matter}\ }\textbf {\bibinfo {volume} {7}},\
  \bibinfo {pages} {8484} (\bibinfo {year} {2011})}\BibitemShut {NoStop}%
\bibitem [{\citenamefont {Ivaneyko}\ \emph {et~al.}(2012)\citenamefont
  {Ivaneyko}, \citenamefont {Toshchevikov}, \citenamefont {Saphiannikova},\
  and\ \citenamefont {Heinrich}}]{Ivaneyko2012}%
  \BibitemOpen
  \bibfield  {author} {\bibinfo {author} {\bibfnamefont {D.}~\bibnamefont
  {Ivaneyko}}, \bibinfo {author} {\bibfnamefont {V.}~\bibnamefont
  {Toshchevikov}}, \bibinfo {author} {\bibfnamefont {M.}~\bibnamefont
  {Saphiannikova}}, \ and\ \bibinfo {author} {\bibfnamefont {G.}~\bibnamefont
  {Heinrich}},\ }\href {\doibase 10.5488/CMP.15.33601} {\bibfield  {journal}
  {\bibinfo  {journal} {Condens. Matter Phys.}\ }\textbf {\bibinfo {volume}
  {15}},\ \bibinfo {pages} {33601} (\bibinfo {year} {2012})}\BibitemShut
  {NoStop}%
\bibitem [{\citenamefont {Fischer}\ and\ \citenamefont
  {Menzel}(2019)}]{Fischer2019}%
  \BibitemOpen
  \bibfield  {author} {\bibinfo {author} {\bibfnamefont {L.}~\bibnamefont
  {Fischer}}\ and\ \bibinfo {author} {\bibfnamefont {A.~M.}\ \bibnamefont
  {Menzel}},\ }\href {\doibase 10.1063/1.5118875} {\bibfield  {journal}
  {\bibinfo  {journal} {J. Chem. Phys.}\ }\textbf {\bibinfo {volume} {151}},\
  \bibinfo {pages} {114906} (\bibinfo {year} {2019})}\BibitemShut {NoStop}%
\bibitem [{\citenamefont {Romeis}, \citenamefont {Toshchevikov},\ and\
  \citenamefont {Saphiannikova}(2019)}]{Romeis2019}%
  \BibitemOpen
  \bibfield  {author} {\bibinfo {author} {\bibfnamefont {D.}~\bibnamefont
  {Romeis}}, \bibinfo {author} {\bibfnamefont {V.}~\bibnamefont
  {Toshchevikov}}, \ and\ \bibinfo {author} {\bibfnamefont {M.}~\bibnamefont
  {Saphiannikova}},\ }\href {\doibase 10.1039/C9SM00226J} {\bibfield  {journal}
  {\bibinfo  {journal} {Soft Matter}\ }\textbf {\bibinfo {volume} {15}},\
  \bibinfo {pages} {3552} (\bibinfo {year} {2019})}\BibitemShut {NoStop}%
\bibitem [{\citenamefont {Wood}\ and\ \citenamefont {Camp}(2011)}]{Wood2011}%
  \BibitemOpen
  \bibfield  {author} {\bibinfo {author} {\bibfnamefont {D.~S.}\ \bibnamefont
  {Wood}}\ and\ \bibinfo {author} {\bibfnamefont {P.~J.}\ \bibnamefont
  {Camp}},\ }\href {\doibase 10.1103/PhysRevE.83.011402} {\bibfield  {journal}
  {\bibinfo  {journal} {Phys. Rev. E}\ }\textbf {\bibinfo {volume} {83}},\
  \bibinfo {pages} {011402} (\bibinfo {year} {2011})}\BibitemShut {NoStop}%
\bibitem [{\citenamefont {Biller}, \citenamefont {Stolbov},\ and\ \citenamefont
  {Raikher}(2014)}]{Biller2014}%
  \BibitemOpen
  \bibfield  {author} {\bibinfo {author} {\bibfnamefont {A.~M.}\ \bibnamefont
  {Biller}}, \bibinfo {author} {\bibfnamefont {O.~V.}\ \bibnamefont {Stolbov}},
  \ and\ \bibinfo {author} {\bibfnamefont {Y.~L.}\ \bibnamefont {Raikher}},\
  }\href {\doibase 10.1063/1.4895980} {\bibfield  {journal} {\bibinfo
  {journal} {J. Appl. Phys.}\ }\textbf {\bibinfo {volume} {116}},\ \bibinfo
  {pages} {114904} (\bibinfo {year} {2014})}\BibitemShut {NoStop}%
\bibitem [{\citenamefont {Biller}, \citenamefont {Stolbov},\ and\ \citenamefont
  {Raikher}(2015)}]{Biller2015}%
  \BibitemOpen
  \bibfield  {author} {\bibinfo {author} {\bibfnamefont {A.~M.}\ \bibnamefont
  {Biller}}, \bibinfo {author} {\bibfnamefont {O.~V.}\ \bibnamefont {Stolbov}},
  \ and\ \bibinfo {author} {\bibfnamefont {Y.~L.}\ \bibnamefont {Raikher}},\
  }\href {\doibase 10.1103/PhysRevE.92.023202} {\bibfield  {journal} {\bibinfo
  {journal} {Phys. Rev. E}\ }\textbf {\bibinfo {volume} {92}},\ \bibinfo
  {pages} {023202} (\bibinfo {year} {2015})}\BibitemShut {NoStop}%
\bibitem [{\citenamefont {Puljiz}\ \emph {et~al.}(2016)\citenamefont {Puljiz},
  \citenamefont {Huang}, \citenamefont {Auernhammer},\ and\ \citenamefont
  {Menzel}}]{Puljiz2016}%
  \BibitemOpen
  \bibfield  {author} {\bibinfo {author} {\bibfnamefont {M.}~\bibnamefont
  {Puljiz}}, \bibinfo {author} {\bibfnamefont {S.}~\bibnamefont {Huang}},
  \bibinfo {author} {\bibfnamefont {G.~K.}\ \bibnamefont {Auernhammer}}, \ and\
  \bibinfo {author} {\bibfnamefont {A.~M.}\ \bibnamefont {Menzel}},\ }\href
  {\doibase 10.1103/PhysRevLett.117.238003} {\bibfield  {journal} {\bibinfo
  {journal} {Phys. Rev. Lett.}\ }\textbf {\bibinfo {volume} {117}},\ \bibinfo
  {pages} {238003} (\bibinfo {year} {2016})}\BibitemShut {NoStop}%
\bibitem [{\citenamefont {Puljiz}\ and\ \citenamefont
  {Menzel}(2017)}]{Puljiz2017}%
  \BibitemOpen
  \bibfield  {author} {\bibinfo {author} {\bibfnamefont {M.}~\bibnamefont
  {Puljiz}}\ and\ \bibinfo {author} {\bibfnamefont {A.~M.}\ \bibnamefont
  {Menzel}},\ }\href {\doibase 10.1103/PhysRevE.95.053002} {\bibfield
  {journal} {\bibinfo  {journal} {Phys. Rev. E}\ }\textbf {\bibinfo {volume}
  {95}},\ \bibinfo {pages} {053002} (\bibinfo {year} {2017})}\BibitemShut
  {NoStop}%
\bibitem [{\citenamefont {Metsch}\ \emph {et~al.}(2016)\citenamefont {Metsch},
  \citenamefont {Kalina}, \citenamefont {Spieler},\ and\ \citenamefont
  {K{\"a}stner}}]{Metsch2016}%
  \BibitemOpen
  \bibfield  {author} {\bibinfo {author} {\bibfnamefont {P.}~\bibnamefont
  {Metsch}}, \bibinfo {author} {\bibfnamefont {K.~A.}\ \bibnamefont {Kalina}},
  \bibinfo {author} {\bibfnamefont {C.}~\bibnamefont {Spieler}}, \ and\
  \bibinfo {author} {\bibfnamefont {M.}~\bibnamefont {K{\"a}stner}},\ }\href
  {\doibase 10.1016/j.commatsci.2016.08.012} {\bibfield  {journal} {\bibinfo
  {journal} {Comput. Mater. Sci.}\ }\textbf {\bibinfo {volume} {124}},\
  \bibinfo {pages} {364 } (\bibinfo {year} {2016})}\BibitemShut {NoStop}%
\bibitem [{\citenamefont {Romeis}\ \emph {et~al.}(2017)\citenamefont {Romeis},
  \citenamefont {Metsch}, \citenamefont {K\"astner},\ and\ \citenamefont
  {Saphiannikova}}]{Romeis2017}%
  \BibitemOpen
  \bibfield  {author} {\bibinfo {author} {\bibfnamefont {D.}~\bibnamefont
  {Romeis}}, \bibinfo {author} {\bibfnamefont {P.}~\bibnamefont {Metsch}},
  \bibinfo {author} {\bibfnamefont {M.}~\bibnamefont {K\"astner}}, \ and\
  \bibinfo {author} {\bibfnamefont {M.}~\bibnamefont {Saphiannikova}},\ }\href
  {\doibase 10.1103/PhysRevE.95.042501} {\bibfield  {journal} {\bibinfo
  {journal} {Phys. Rev. E}\ }\textbf {\bibinfo {volume} {95}},\ \bibinfo
  {pages} {042501} (\bibinfo {year} {2017})}\BibitemShut {NoStop}%
\bibitem [{\citenamefont {Pessot}\ \emph {et~al.}(2014)\citenamefont {Pessot},
  \citenamefont {Cremer}, \citenamefont {Borin}, \citenamefont {Odenbach},
  \citenamefont {L{\"o}wen},\ and\ \citenamefont {Menzel}}]{Pessot2014}%
  \BibitemOpen
  \bibfield  {author} {\bibinfo {author} {\bibfnamefont {G.}~\bibnamefont
  {Pessot}}, \bibinfo {author} {\bibfnamefont {P.}~\bibnamefont {Cremer}},
  \bibinfo {author} {\bibfnamefont {D.~Y.}\ \bibnamefont {Borin}}, \bibinfo
  {author} {\bibfnamefont {S.}~\bibnamefont {Odenbach}}, \bibinfo {author}
  {\bibfnamefont {H.}~\bibnamefont {L{\"o}wen}}, \ and\ \bibinfo {author}
  {\bibfnamefont {A.~M.}\ \bibnamefont {Menzel}},\ }\href {\doibase
  10.1063/1.4896147} {\bibfield  {journal} {\bibinfo  {journal} {J. Chem.
  Phys.}\ }\textbf {\bibinfo {volume} {141}},\ \bibinfo {pages} {015005}
  (\bibinfo {year} {2014})}\BibitemShut {NoStop}%
\bibitem [{\citenamefont {Pessot}, \citenamefont {L{\"o}wen},\ and\
  \citenamefont {Menzel}(2016)}]{Pessot2016}%
  \BibitemOpen
  \bibfield  {author} {\bibinfo {author} {\bibfnamefont {G.}~\bibnamefont
  {Pessot}}, \bibinfo {author} {\bibfnamefont {H.}~\bibnamefont {L{\"o}wen}}, \
  and\ \bibinfo {author} {\bibfnamefont {A.~M.}\ \bibnamefont {Menzel}},\
  }\href {\doibase 10.1063/1.4962365} {\bibfield  {journal} {\bibinfo
  {journal} {J. Chem. Phys.}\ }\textbf {\bibinfo {volume} {145}},\ \bibinfo
  {pages} {104904} (\bibinfo {year} {2016})}\BibitemShut {NoStop}%
\bibitem [{\citenamefont {Menzel}(2019)}]{Menzel2019}%
  \BibitemOpen
  \bibfield  {author} {\bibinfo {author} {\bibfnamefont {A.~M.}\ \bibnamefont
  {Menzel}},\ }\href {\doibase 10.1007/s00419-018-1413-7} {\bibfield  {journal}
  {\bibinfo  {journal} {Arch. Appl. Mech.}\ }\textbf {\bibinfo {volume} {89}},\
  \bibinfo {pages} {17} (\bibinfo {year} {2019})}\BibitemShut {NoStop}%
\bibitem [{\citenamefont {Menzel}(2014)}]{Menzel2014}%
  \BibitemOpen
  \bibfield  {author} {\bibinfo {author} {\bibfnamefont {A.~M.}\ \bibnamefont
  {Menzel}},\ }\href {\doibase 10.1063/1.4901275} {\bibfield  {journal}
  {\bibinfo  {journal} {J. Chem. Phys.}\ }\textbf {\bibinfo {volume} {141}},\
  \bibinfo {pages} {194907} (\bibinfo {year} {2014})}\BibitemShut {NoStop}%
\bibitem [{\citenamefont {Goh}\ \emph {et~al.}(2019)\citenamefont {Goh},
  \citenamefont {Wittmann}, \citenamefont {Menzel},\ and\ \citenamefont
  {L\"owen}}]{Goh2019}%
  \BibitemOpen
  \bibfield  {author} {\bibinfo {author} {\bibfnamefont {S.}~\bibnamefont
  {Goh}}, \bibinfo {author} {\bibfnamefont {R.}~\bibnamefont {Wittmann}},
  \bibinfo {author} {\bibfnamefont {A.~M.}\ \bibnamefont {Menzel}}, \ and\
  \bibinfo {author} {\bibfnamefont {H.}~\bibnamefont {L\"owen}},\ }\href
  {\doibase 10.1103/PhysRevE.100.012605} {\bibfield  {journal} {\bibinfo
  {journal} {Phys. Rev. E}\ }\textbf {\bibinfo {volume} {100}},\ \bibinfo
  {pages} {012605} (\bibinfo {year} {2019})}\BibitemShut {NoStop}%
\bibitem [{\citenamefont {Evans}(1979)}]{Evans1979}%
  \BibitemOpen
  \bibfield  {author} {\bibinfo {author} {\bibfnamefont {R.}~\bibnamefont
  {Evans}},\ }\href {\doibase 10.1080/00018737900101365} {\bibfield  {journal}
  {\bibinfo  {journal} {Adv. Phys.}\ }\textbf {\bibinfo {volume} {28}},\
  \bibinfo {pages} {143} (\bibinfo {year} {1979})}\BibitemShut {NoStop}%
\bibitem [{\citenamefont {L{\"o}wen}(2002)}]{Lowen2002}%
  \BibitemOpen
  \bibfield  {author} {\bibinfo {author} {\bibfnamefont {H.}~\bibnamefont
  {L{\"o}wen}},\ }\href {\doibase 10.1088/0953-8984/14/46/301} {\bibfield
  {journal} {\bibinfo  {journal} {J. Phys.: Condens. Matter}\ }\textbf
  {\bibinfo {volume} {14}},\ \bibinfo {pages} {11897} (\bibinfo {year}
  {2002})}\BibitemShut {NoStop}%
\bibitem [{\citenamefont {Oxtoby}(2002)}]{Oxtoby2002}%
  \BibitemOpen
  \bibfield  {author} {\bibinfo {author} {\bibfnamefont {D.~W.}\ \bibnamefont
  {Oxtoby}},\ }\href {\doibase 10.1146/annurev.matsci.32.100401.103425}
  {\bibfield  {journal} {\bibinfo  {journal} {Annu. Rev. Mater. Res.}\ }\textbf
  {\bibinfo {volume} {32}},\ \bibinfo {pages} {39} (\bibinfo {year}
  {2002})}\BibitemShut {NoStop}%
\bibitem [{\citenamefont {Evans}\ \emph {et~al.}(2016)\citenamefont {Evans},
  \citenamefont {Oettel}, \citenamefont {Roth},\ and\ \citenamefont
  {Kahl}}]{Evans2016}%
  \BibitemOpen
  \bibfield  {author} {\bibinfo {author} {\bibfnamefont {R.}~\bibnamefont
  {Evans}}, \bibinfo {author} {\bibfnamefont {M.}~\bibnamefont {Oettel}},
  \bibinfo {author} {\bibfnamefont {R.}~\bibnamefont {Roth}}, \ and\ \bibinfo
  {author} {\bibfnamefont {G.}~\bibnamefont {Kahl}},\ }\href {\doibase
  10.1088/0953-8984/28/24/240401} {\bibfield  {journal} {\bibinfo  {journal}
  {J. Phys.: Condens. Matter}\ }\textbf {\bibinfo {volume} {28}},\ \bibinfo
  {pages} {240401} (\bibinfo {year} {2016})}\BibitemShut {NoStop}%
\bibitem [{\citenamefont {Dauxois}\ \emph {et~al.}(2002)\citenamefont
  {Dauxois}, \citenamefont {Ruffo}, \citenamefont {Arimondo},\ and\
  \citenamefont {Wilkens}}]{Dauxois2002}%
  \BibitemOpen
  \bibfield  {author} {\bibinfo {author} {\bibfnamefont {T.}~\bibnamefont
  {Dauxois}}, \bibinfo {author} {\bibfnamefont {S.}~\bibnamefont {Ruffo}},
  \bibinfo {author} {\bibfnamefont {E.}~\bibnamefont {Arimondo}}, \ and\
  \bibinfo {author} {\bibfnamefont {M.}~\bibnamefont {Wilkens}},\ }in\
  \href@noop {} {\emph {\bibinfo {booktitle} {Dynamics and Thermodynamics of
  Systems with Long-Range Interactions}}}\ (\bibinfo  {publisher} {Springer},\
  \bibinfo {year} {2002})\ pp.\ \bibinfo {pages} {1--19}\BibitemShut {NoStop}%
\bibitem [{\citenamefont {Arnold}\ and\ \citenamefont
  {Holm}(2005)}]{Arnold2005}%
  \BibitemOpen
  \bibfield  {author} {\bibinfo {author} {\bibfnamefont {A.}~\bibnamefont
  {Arnold}}\ and\ \bibinfo {author} {\bibfnamefont {C.}~\bibnamefont {Holm}},\
  }\enquote {\bibinfo {title} {Efficient methods to compute long-range
  interactions for soft matter systems},}\ in\ \href {\doibase 10.1007/b136793}
  {\emph {\bibinfo {booktitle} {Advanced Computer Simulation Approaches for
  Soft Matter Sciences II}}},\ \bibinfo {editor} {edited by\ \bibinfo {editor}
  {\bibfnamefont {C.}~\bibnamefont {Holm}}\ and\ \bibinfo {editor}
  {\bibfnamefont {K.}~\bibnamefont {Kremer}}}\ (\bibinfo  {publisher}
  {Springer},\ \bibinfo {address} {Berlin Heidelberg},\ \bibinfo {year}
  {2005})\ pp.\ \bibinfo {pages} {59--109}\BibitemShut {NoStop}%
\bibitem [{\citenamefont {Allen}\ and\ \citenamefont
  {Tildesley}(2017)}]{Allen2017}%
  \BibitemOpen
  \bibfield  {author} {\bibinfo {author} {\bibfnamefont {M.~P.}\ \bibnamefont
  {Allen}}\ and\ \bibinfo {author} {\bibfnamefont {D.~J.}\ \bibnamefont
  {Tildesley}},\ }\href@noop {} {\emph {\bibinfo {title} {Computer Simulation
  of Liquids}}}\ (\bibinfo  {publisher} {Oxford University Press},\ \bibinfo
  {year} {2017})\BibitemShut {NoStop}%
\bibitem [{\citenamefont {Peroukidis}\ and\ \citenamefont
  {Klapp}(2016)}]{Peroukidis2016}%
  \BibitemOpen
  \bibfield  {author} {\bibinfo {author} {\bibfnamefont {S.~D.}\ \bibnamefont
  {Peroukidis}}\ and\ \bibinfo {author} {\bibfnamefont {S.~H.~L.}\ \bibnamefont
  {Klapp}},\ }\href {\doibase 10.1039/C6SM01264G} {\bibfield  {journal}
  {\bibinfo  {journal} {Soft Matter}\ }\textbf {\bibinfo {volume} {12}},\
  \bibinfo {pages} {6841} (\bibinfo {year} {2016})}\BibitemShut {NoStop}%
\bibitem [{\citenamefont {Siboni}\ \emph {et~al.}(2020)\citenamefont {Siboni},
  \citenamefont {Shrivastav}, \citenamefont {Peroukidis},\ and\ \citenamefont
  {Klapp}}]{Siboni2020}%
  \BibitemOpen
  \bibfield  {author} {\bibinfo {author} {\bibfnamefont {N.~H.}\ \bibnamefont
  {Siboni}}, \bibinfo {author} {\bibfnamefont {G.~P.}\ \bibnamefont
  {Shrivastav}}, \bibinfo {author} {\bibfnamefont {S.~D.}\ \bibnamefont
  {Peroukidis}}, \ and\ \bibinfo {author} {\bibfnamefont {S.~H.~L.}\
  \bibnamefont {Klapp}},\ }\href {\doibase 10.1515/psr-2019-0108} {\bibfield
  {journal} {\bibinfo  {journal} {Phys. Sci. Rev.}\ }\textbf {\bibinfo {volume}
  {2019}},\ \bibinfo {pages} {20190108} (\bibinfo {year} {2020})}\BibitemShut
  {NoStop}%
\bibitem [{\citenamefont {Groh}\ and\ \citenamefont
  {Dietrich}(1994{\natexlab{a}})}]{Groh1994prl}%
  \BibitemOpen
  \bibfield  {author} {\bibinfo {author} {\bibfnamefont {B.}~\bibnamefont
  {Groh}}\ and\ \bibinfo {author} {\bibfnamefont {S.}~\bibnamefont
  {Dietrich}},\ }\href {\doibase 10.1103/PhysRevLett.72.2422} {\bibfield
  {journal} {\bibinfo  {journal} {Phys. Rev. Lett.}\ }\textbf {\bibinfo
  {volume} {72}},\ \bibinfo {pages} {2422} (\bibinfo {year}
  {1994}{\natexlab{a}})}\BibitemShut {NoStop}%
\bibitem [{\citenamefont {Groh}\ and\ \citenamefont
  {Dietrich}(1994{\natexlab{b}})}]{Groh1994pre}%
  \BibitemOpen
  \bibfield  {author} {\bibinfo {author} {\bibfnamefont {B.}~\bibnamefont
  {Groh}}\ and\ \bibinfo {author} {\bibfnamefont {S.}~\bibnamefont
  {Dietrich}},\ }\href {\doibase 10.1103/PhysRevE.50.3814} {\bibfield
  {journal} {\bibinfo  {journal} {Phys. Rev. E}\ }\textbf {\bibinfo {volume}
  {50}},\ \bibinfo {pages} {3814} (\bibinfo {year}
  {1994}{\natexlab{b}})}\BibitemShut {NoStop}%
\bibitem [{\citenamefont {Ivaneyko}\ \emph {et~al.}(2014)\citenamefont
  {Ivaneyko}, \citenamefont {Toshchevikov}, \citenamefont {Saphiannikova},\
  and\ \citenamefont {Heinrich}}]{Ivaneyko2014}%
  \BibitemOpen
  \bibfield  {author} {\bibinfo {author} {\bibfnamefont {D.}~\bibnamefont
  {Ivaneyko}}, \bibinfo {author} {\bibfnamefont {V.}~\bibnamefont
  {Toshchevikov}}, \bibinfo {author} {\bibfnamefont {M.}~\bibnamefont
  {Saphiannikova}}, \ and\ \bibinfo {author} {\bibfnamefont {G.}~\bibnamefont
  {Heinrich}},\ }\href {\doibase 10.1039/C3SM52440J} {\bibfield  {journal}
  {\bibinfo  {journal} {Soft Matter}\ }\textbf {\bibinfo {volume} {10}},\
  \bibinfo {pages} {2213} (\bibinfo {year} {2014})}\BibitemShut {NoStop}%
\bibitem [{\citenamefont {de~Gennes}(1980)}]{deGennes1980}%
  \BibitemOpen
  \bibfield  {author} {\bibinfo {author} {\bibfnamefont {P.~G.}\ \bibnamefont
  {de~Gennes}},\ }\enquote {\bibinfo {title} {Weak nematic gels},}\ in\
  \href@noop {} {\emph {\bibinfo {booktitle} {Liquid Crystals of One-and
  Two-Dimensional Order}}}\ (\bibinfo  {publisher} {Springer},\ \bibinfo {year}
  {1980})\ pp.\ \bibinfo {pages} {231--237}\BibitemShut {NoStop}%
\bibitem [{\citenamefont {Brand}\ and\ \citenamefont
  {Pleiner}(1994)}]{Brand1994}%
  \BibitemOpen
  \bibfield  {author} {\bibinfo {author} {\bibfnamefont {H.~R.}\ \bibnamefont
  {Brand}}\ and\ \bibinfo {author} {\bibfnamefont {H.}~\bibnamefont
  {Pleiner}},\ }\href {\doibase 10.1016/0378-4371(94)00060-3} {\bibfield
  {journal} {\bibinfo  {journal} {Physica A}\ }\textbf {\bibinfo {volume}
  {208}},\ \bibinfo {pages} {359} (\bibinfo {year} {1994})}\BibitemShut
  {NoStop}%
\bibitem [{\citenamefont {Menzel}, \citenamefont {Pleiner},\ and\ \citenamefont
  {Brand}(2007)}]{Menzel2007}%
  \BibitemOpen
  \bibfield  {author} {\bibinfo {author} {\bibfnamefont {A.~M.}\ \bibnamefont
  {Menzel}}, \bibinfo {author} {\bibfnamefont {H.}~\bibnamefont {Pleiner}}, \
  and\ \bibinfo {author} {\bibfnamefont {H.~R.}\ \bibnamefont {Brand}},\ }\href
  {\doibase 10.1063/1.2742383} {\bibfield  {journal} {\bibinfo  {journal} {J.
  Chem. Phys.}\ }\textbf {\bibinfo {volume} {126}},\ \bibinfo {pages} {234901}
  (\bibinfo {year} {2007})}\BibitemShut {NoStop}%
\bibitem [{\citenamefont {Menzel}, \citenamefont {Pleiner},\ and\ \citenamefont
  {Brand}(2009)}]{Menzel2009}%
  \BibitemOpen
  \bibfield  {author} {\bibinfo {author} {\bibfnamefont {A.~M.}\ \bibnamefont
  {Menzel}}, \bibinfo {author} {\bibfnamefont {H.}~\bibnamefont {Pleiner}}, \
  and\ \bibinfo {author} {\bibfnamefont {H.~R.}\ \bibnamefont {Brand}},\ }\href
  {\doibase 10.1140/epje/i2009-10535-2} {\bibfield  {journal} {\bibinfo
  {journal} {Eur. Phys. J. E}\ }\textbf {\bibinfo {volume} {30}},\ \bibinfo
  {pages} {371} (\bibinfo {year} {2009})}\BibitemShut {NoStop}%
\bibitem [{\citenamefont {Filipcsei}\ and\ \citenamefont
  {Zr{\'{\i}}nyi}(2010)}]{Filipcsei2010}%
  \BibitemOpen
  \bibfield  {author} {\bibinfo {author} {\bibfnamefont {G.}~\bibnamefont
  {Filipcsei}}\ and\ \bibinfo {author} {\bibfnamefont {M.}~\bibnamefont
  {Zr{\'{\i}}nyi}},\ }\href {\doibase 10.1088/0953-8984/22/27/276001}
  {\bibfield  {journal} {\bibinfo  {journal} {J. Phys.: Condens. Matter}\
  }\textbf {\bibinfo {volume} {22}},\ \bibinfo {pages} {276001} (\bibinfo
  {year} {2010})}\BibitemShut {NoStop}%
\bibitem [{\citenamefont {Annunziata}, \citenamefont {Menzel},\ and\
  \citenamefont {L\"owen}(2013)}]{Annunziata2013}%
  \BibitemOpen
  \bibfield  {author} {\bibinfo {author} {\bibfnamefont {M.~A.}\ \bibnamefont
  {Annunziata}}, \bibinfo {author} {\bibfnamefont {A.~M.}\ \bibnamefont
  {Menzel}}, \ and\ \bibinfo {author} {\bibfnamefont {H.}~\bibnamefont
  {L\"owen}},\ }\href {\doibase 10.1063/1.4807003} {\bibfield  {journal}
  {\bibinfo  {journal} {J. Chem. Phys.}\ }\textbf {\bibinfo {volume} {138}},\
  \bibinfo {pages} {204906} (\bibinfo {year} {2013})}\BibitemShut {NoStop}%
\bibitem [{\citenamefont {S{\'a}nchez}\ \emph {et~al.}(2019)\citenamefont
  {S{\'a}nchez}, \citenamefont {Stolbov}, \citenamefont {Kantorovich},\ and\
  \citenamefont {Raikher}}]{Sanchez2019}%
  \BibitemOpen
  \bibfield  {author} {\bibinfo {author} {\bibfnamefont {P.~A.}\ \bibnamefont
  {S{\'a}nchez}}, \bibinfo {author} {\bibfnamefont {O.~V.}\ \bibnamefont
  {Stolbov}}, \bibinfo {author} {\bibfnamefont {S.~S.}\ \bibnamefont
  {Kantorovich}}, \ and\ \bibinfo {author} {\bibfnamefont {Y.~L.}\ \bibnamefont
  {Raikher}},\ }\href {\doibase 10.1039/C9SM00827F} {\bibfield  {journal}
  {\bibinfo  {journal} {Soft Matter}\ }\textbf {\bibinfo {volume} {15}},\
  \bibinfo {pages} {7145} (\bibinfo {year} {2019})}\BibitemShut {NoStop}%
\bibitem [{\citenamefont {Vaganov}\ \emph {et~al.}(2020)\citenamefont
  {Vaganov}, \citenamefont {Borin}, \citenamefont {Odenbach},\ and\
  \citenamefont {Raikher}}]{Vaganov2020}%
  \BibitemOpen
  \bibfield  {author} {\bibinfo {author} {\bibfnamefont {M.~V.}\ \bibnamefont
  {Vaganov}}, \bibinfo {author} {\bibfnamefont {D.~Y.}\ \bibnamefont {Borin}},
  \bibinfo {author} {\bibfnamefont {S.}~\bibnamefont {Odenbach}}, \ and\
  \bibinfo {author} {\bibfnamefont {Y.~L.}\ \bibnamefont {Raikher}},\ }\href
  {\doibase 10.1016/j.jmmm.2019.166249} {\bibfield  {journal} {\bibinfo
  {journal} {J. Magn. Magn. Mater.}\ }\textbf {\bibinfo {volume} {499}},\
  \bibinfo {pages} {166249} (\bibinfo {year} {2020})}\BibitemShut {NoStop}%
\bibitem [{\citenamefont {Goh}, \citenamefont {L\"owen},\ and\ \citenamefont
  {Menzel}(2022)}]{Goh2022}%
  \BibitemOpen
  \bibfield  {author} {\bibinfo {author} {\bibfnamefont {S.}~\bibnamefont
  {Goh}}, \bibinfo {author} {\bibfnamefont {H.}~\bibnamefont {L\"owen}}, \ and\
  \bibinfo {author} {\bibfnamefont {A.~M.}\ \bibnamefont {Menzel}},\ }\href
  {\doibase 10.1103/PhysRevB.106.L100101} {\bibfield  {journal} {\bibinfo
  {journal} {Phys. Rev. B}\ }\textbf {\bibinfo {volume} {106}},\ \bibinfo
  {pages} {L100101} (\bibinfo {year} {2022})}\BibitemShut {NoStop}%
\bibitem [{\citenamefont {Mehrabadi}\ and\ \citenamefont
  {Cowin}(1990)}]{Mehrabadi1990}%
  \BibitemOpen
  \bibfield  {author} {\bibinfo {author} {\bibfnamefont {M.~M.}\ \bibnamefont
  {Mehrabadi}}\ and\ \bibinfo {author} {\bibfnamefont {S.~C.}\ \bibnamefont
  {Cowin}},\ }\href {\doibase 10.1093/qjmam/43.1.15} {\bibfield  {journal}
  {\bibinfo  {journal} {Q. J. Mech. Appl. Math.}\ }\textbf {\bibinfo {volume}
  {43}},\ \bibinfo {pages} {15} (\bibinfo {year} {1990})}\BibitemShut {NoStop}%
\bibitem [{\citenamefont {Ma{\'{z}}dziarz}(2019)}]{Mazdziarz2019}%
  \BibitemOpen
  \bibfield  {author} {\bibinfo {author} {\bibfnamefont {M.}~\bibnamefont
  {Ma{\'{z}}dziarz}},\ }\href {\doibase 10.1088/2053-1583/ab2ef3} {\bibfield
  {journal} {\bibinfo  {journal} {2D Mater.}\ }\textbf {\bibinfo {volume}
  {6}},\ \bibinfo {pages} {048001} (\bibinfo {year} {2019})}\BibitemShut
  {NoStop}%
\bibitem [{\citenamefont {Brugger}(1965)}]{Brugger1965}%
  \BibitemOpen
  \bibfield  {author} {\bibinfo {author} {\bibfnamefont {K.}~\bibnamefont
  {Brugger}},\ }\href {\doibase 10.1063/1.1714215} {\bibfield  {journal}
  {\bibinfo  {journal} {J. Appl. Phys.}\ }\textbf {\bibinfo {volume} {36}},\
  \bibinfo {pages} {759} (\bibinfo {year} {1965})}\BibitemShut {NoStop}%
\bibitem [{\citenamefont {Clayton}(2010)}]{Clayton2010}%
  \BibitemOpen
  \bibfield  {author} {\bibinfo {author} {\bibfnamefont {J.~D.}\ \bibnamefont
  {Clayton}},\ }\href@noop {} {\emph {\bibinfo {title} {Nonlinear mechanics of
  crystals}}},\ Vol.\ \bibinfo {volume} {177}\ (\bibinfo  {publisher} {Springer
  Science \& Business Media},\ \bibinfo {year} {2010})\ Chap.\ \bibinfo
  {chapter} {Appendix A}\BibitemShut {NoStop}%
\bibitem [{\citenamefont {Oettel}\ \emph {et~al.}(2010)\citenamefont {Oettel},
  \citenamefont {G\"orig}, \citenamefont {H\"artel}, \citenamefont {L\"owen},
  \citenamefont {Radu},\ and\ \citenamefont {Schilling}}]{Oettel2010}%
  \BibitemOpen
  \bibfield  {author} {\bibinfo {author} {\bibfnamefont {M.}~\bibnamefont
  {Oettel}}, \bibinfo {author} {\bibfnamefont {S.}~\bibnamefont {G\"orig}},
  \bibinfo {author} {\bibfnamefont {A.}~\bibnamefont {H\"artel}}, \bibinfo
  {author} {\bibfnamefont {H.}~\bibnamefont {L\"owen}}, \bibinfo {author}
  {\bibfnamefont {M.}~\bibnamefont {Radu}}, \ and\ \bibinfo {author}
  {\bibfnamefont {T.}~\bibnamefont {Schilling}},\ }\href {\doibase
  10.1103/PhysRevE.82.051404} {\bibfield  {journal} {\bibinfo  {journal} {Phys.
  Rev. E}\ }\textbf {\bibinfo {volume} {82}},\ \bibinfo {pages} {051404}
  (\bibinfo {year} {2010})}\BibitemShut {NoStop}%
\bibitem [{\citenamefont {Hansen-Goos}\ and\ \citenamefont
  {Roth}(2006)}]{HansenGoos2006}%
  \BibitemOpen
  \bibfield  {author} {\bibinfo {author} {\bibfnamefont {H.}~\bibnamefont
  {Hansen-Goos}}\ and\ \bibinfo {author} {\bibfnamefont {R.}~\bibnamefont
  {Roth}},\ }\href {\doibase 10.1088/0953-8984/18/37/002} {\bibfield  {journal}
  {\bibinfo  {journal} {J. Phys.: Condens. Matter}\ }\textbf {\bibinfo {volume}
  {18}},\ \bibinfo {pages} {8413} (\bibinfo {year} {2006})}\BibitemShut
  {NoStop}%
\bibitem [{\citenamefont {Tarazona}(2000)}]{Tarazona2000}%
  \BibitemOpen
  \bibfield  {author} {\bibinfo {author} {\bibfnamefont {P.}~\bibnamefont
  {Tarazona}},\ }\href {\doibase 10.1103/PhysRevLett.84.694} {\bibfield
  {journal} {\bibinfo  {journal} {Phys. Rev. Lett.}\ }\textbf {\bibinfo
  {volume} {84}},\ \bibinfo {pages} {694} (\bibinfo {year} {2000})}\BibitemShut
  {NoStop}%
\bibitem [{\citenamefont {Roth}(2010)}]{Roth2010}%
  \BibitemOpen
  \bibfield  {author} {\bibinfo {author} {\bibfnamefont {R.}~\bibnamefont
  {Roth}},\ }\href {\doibase 10.1088/0953-8984/22/6/063102} {\bibfield
  {journal} {\bibinfo  {journal} {J. Phys.: Condens. Matter}\ }\textbf
  {\bibinfo {volume} {22}},\ \bibinfo {pages} {063102} (\bibinfo {year}
  {2010})}\BibitemShut {NoStop}%
\bibitem [{\citenamefont {Klapp}(2005)}]{Klapp2005}%
  \BibitemOpen
  \bibfield  {author} {\bibinfo {author} {\bibfnamefont {S.~H.~L.}\
  \bibnamefont {Klapp}},\ }\href {\doibase 10.1088/0953-8984/17/15/r02}
  {\bibfield  {journal} {\bibinfo  {journal} {J. Phys.: Condens. Matter}\
  }\textbf {\bibinfo {volume} {17}},\ \bibinfo {pages} {R525} (\bibinfo {year}
  {2005})}\BibitemShut {NoStop}%
\bibitem [{\citenamefont {Ramakrishnan}\ and\ \citenamefont
  {Yussouff}(1979)}]{Ramakrishnan1979}%
  \BibitemOpen
  \bibfield  {author} {\bibinfo {author} {\bibfnamefont {T.~V.}\ \bibnamefont
  {Ramakrishnan}}\ and\ \bibinfo {author} {\bibfnamefont {M.}~\bibnamefont
  {Yussouff}},\ }\href {\doibase 10.1103/PhysRevB.19.2775} {\bibfield
  {journal} {\bibinfo  {journal} {Phys. Rev. B}\ }\textbf {\bibinfo {volume}
  {19}},\ \bibinfo {pages} {2775} (\bibinfo {year} {1979})}\BibitemShut
  {NoStop}%
\bibitem [{\citenamefont {Curtin}\ and\ \citenamefont
  {Ashcroft}(1985)}]{Curtin1985}%
  \BibitemOpen
  \bibfield  {author} {\bibinfo {author} {\bibfnamefont {W.~A.}\ \bibnamefont
  {Curtin}}\ and\ \bibinfo {author} {\bibfnamefont {N.~W.}\ \bibnamefont
  {Ashcroft}},\ }\href {\doibase 10.1103/PhysRevA.32.2909} {\bibfield
  {journal} {\bibinfo  {journal} {Phys. Rev. A}\ }\textbf {\bibinfo {volume}
  {32}},\ \bibinfo {pages} {2909} (\bibinfo {year} {1985})}\BibitemShut
  {NoStop}%
\bibitem [{\citenamefont {Denton}\ and\ \citenamefont
  {Ashcroft}(1989)}]{Denton1989}%
  \BibitemOpen
  \bibfield  {author} {\bibinfo {author} {\bibfnamefont {A.~R.}\ \bibnamefont
  {Denton}}\ and\ \bibinfo {author} {\bibfnamefont {N.~W.}\ \bibnamefont
  {Ashcroft}},\ }\href {\doibase 10.1103/PhysRevA.39.4701} {\bibfield
  {journal} {\bibinfo  {journal} {Phys. Rev. A}\ }\textbf {\bibinfo {volume}
  {39}},\ \bibinfo {pages} {4701} (\bibinfo {year} {1989})}\BibitemShut
  {NoStop}%
\bibitem [{\citenamefont {Rosenfeld}(1989)}]{Rosenfeld1989}%
  \BibitemOpen
  \bibfield  {author} {\bibinfo {author} {\bibfnamefont {Y.}~\bibnamefont
  {Rosenfeld}},\ }\href {\doibase 10.1103/PhysRevLett.63.980} {\bibfield
  {journal} {\bibinfo  {journal} {Phys. Rev. Lett.}\ }\textbf {\bibinfo
  {volume} {63}},\ \bibinfo {pages} {980} (\bibinfo {year} {1989})}\BibitemShut
  {NoStop}%
\bibitem [{\citenamefont {Ohnesorge}, \citenamefont {L{\"o}wen},\ and\
  \citenamefont {Wagner}(1993)}]{Ohnesorge1993}%
  \BibitemOpen
  \bibfield  {author} {\bibinfo {author} {\bibfnamefont {R.}~\bibnamefont
  {Ohnesorge}}, \bibinfo {author} {\bibfnamefont {H.}~\bibnamefont
  {L{\"o}wen}}, \ and\ \bibinfo {author} {\bibfnamefont {H.}~\bibnamefont
  {Wagner}},\ }\href {\doibase 10.1209/0295-5075/22/4/002} {\bibfield
  {journal} {\bibinfo  {journal} {Europhys. Lett.}\ }\textbf {\bibinfo {volume}
  {22}},\ \bibinfo {pages} {245} (\bibinfo {year} {1993})}\BibitemShut
  {NoStop}%
\bibitem [{\citenamefont {Goh}, \citenamefont {Menzel},\ and\ \citenamefont
  {L{\"o}wen}(2018)}]{Goh2018}%
  \BibitemOpen
  \bibfield  {author} {\bibinfo {author} {\bibfnamefont {S.}~\bibnamefont
  {Goh}}, \bibinfo {author} {\bibfnamefont {A.~M.}\ \bibnamefont {Menzel}}, \
  and\ \bibinfo {author} {\bibfnamefont {H.}~\bibnamefont {L{\"o}wen}},\ }\href
  {\doibase 10.1039/C8CP01395K} {\bibfield  {journal} {\bibinfo  {journal}
  {Phys. Chem. Chem. Phys.}\ }\textbf {\bibinfo {volume} {20}},\ \bibinfo
  {pages} {15037} (\bibinfo {year} {2018})}\BibitemShut {NoStop}%
\bibitem [{\citenamefont {Allahyarov}, \citenamefont {L{\"o}wen},\ and\
  \citenamefont {Zhu}(2015)}]{Allahyarov2015}%
  \BibitemOpen
  \bibfield  {author} {\bibinfo {author} {\bibfnamefont {E.}~\bibnamefont
  {Allahyarov}}, \bibinfo {author} {\bibfnamefont {H.}~\bibnamefont
  {L{\"o}wen}}, \ and\ \bibinfo {author} {\bibfnamefont {L.}~\bibnamefont
  {Zhu}},\ }\href {\doibase 10.1039/C5CP05522A} {\bibfield  {journal} {\bibinfo
   {journal} {Phys. Chem. Chem. Phys.}\ }\textbf {\bibinfo {volume} {17}},\
  \bibinfo {pages} {32479} (\bibinfo {year} {2015})}\BibitemShut {NoStop}%
\bibitem [{\citenamefont {Jäger}\ \emph {et~al.}(2022)\citenamefont {Jäger},
  \citenamefont {Fischer}, \citenamefont {Lutz},\ and\ \citenamefont
  {Menzel}}]{Jaeger2022}%
  \BibitemOpen
  \bibfield  {author} {\bibinfo {author} {\bibfnamefont {G.~J.~L.}\
  \bibnamefont {Jäger}}, \bibinfo {author} {\bibfnamefont {L.}~\bibnamefont
  {Fischer}}, \bibinfo {author} {\bibfnamefont {T.}~\bibnamefont {Lutz}}, \
  and\ \bibinfo {author} {\bibfnamefont {A.~M.}\ \bibnamefont {Menzel}},\
  }\href {\doibase 10.1088/1361-648X/ac98e8} {\bibfield  {journal} {\bibinfo
  {journal} {J. Phys.: Condens. Matter}\ }\textbf {\bibinfo {volume} {34}},\
  \bibinfo {pages} {485101} (\bibinfo {year} {2022})}\BibitemShut {NoStop}%
\bibitem [{\citenamefont {Queralto~Gratacos}(2010)}]{Queralto2010}%
  \BibitemOpen
  \bibfield  {author} {\bibinfo {author} {\bibfnamefont {N.}~\bibnamefont
  {Queralto~Gratacos}},\ }\emph {\bibinfo {title} {Functional Hydrogels:
  Ferrogel Thin Films}},\ \href@noop {} {Ph.D. thesis},\ \bibinfo  {school}
  {Johannes Gutenberg-Universit{\"a}t Mainz} (\bibinfo {year}
  {2010})\BibitemShut {NoStop}%
\bibitem [{\citenamefont {van Teeffelen}, \citenamefont {L{\"o}wen},\ and\
  \citenamefont {Likos}(2008)}]{vanTeeffelen2008}%
  \BibitemOpen
  \bibfield  {author} {\bibinfo {author} {\bibfnamefont {S.}~\bibnamefont {van
  Teeffelen}}, \bibinfo {author} {\bibfnamefont {H.}~\bibnamefont {L{\"o}wen}},
  \ and\ \bibinfo {author} {\bibfnamefont {C.~N.}\ \bibnamefont {Likos}},\
  }\href {\doibase 10.1088/0953-8984/20/40/404217} {\bibfield  {journal}
  {\bibinfo  {journal} {J. Phys.: Condens. Matter}\ }\textbf {\bibinfo {volume}
  {20}},\ \bibinfo {pages} {404217} (\bibinfo {year} {2008})}\BibitemShut
  {NoStop}%
\bibitem [{\citenamefont {Fischer}\ and\ \citenamefont
  {Menzel}(2020)}]{Fischer2020}%
  \BibitemOpen
  \bibfield  {author} {\bibinfo {author} {\bibfnamefont {L.}~\bibnamefont
  {Fischer}}\ and\ \bibinfo {author} {\bibfnamefont {A.~M.}\ \bibnamefont
  {Menzel}},\ }\href {\doibase 10.1088/1361-665x/abc148} {\bibfield  {journal}
  {\bibinfo  {journal} {Smart Mater. Struct.}\ }\textbf {\bibinfo {volume}
  {30}},\ \bibinfo {pages} {014003} (\bibinfo {year} {2020})}\BibitemShut
  {NoStop}%
\bibitem [{\citenamefont {Reich}\ and\ \citenamefont
  {Schmidt}(2004)}]{Reich2004}%
  \BibitemOpen
  \bibfield  {author} {\bibinfo {author} {\bibfnamefont {H.}~\bibnamefont
  {Reich}}\ and\ \bibinfo {author} {\bibfnamefont {M.}~\bibnamefont
  {Schmidt}},\ }\href {\doibase 10.1023/B:JOSS.0000041752.55138.0a} {\bibfield
  {journal} {\bibinfo  {journal} {J. Stat. Phys.}\ }\textbf {\bibinfo {volume}
  {116}},\ \bibinfo {pages} {1683} (\bibinfo {year} {2004})}\BibitemShut
  {NoStop}%
\bibitem [{\citenamefont {Schmidt}(2005)}]{Schmidt2005}%
  \BibitemOpen
  \bibfield  {author} {\bibinfo {author} {\bibfnamefont {M.}~\bibnamefont
  {Schmidt}},\ }\href {\doibase 10.1088/0953-8984/17/45/037} {\bibfield
  {journal} {\bibinfo  {journal} {J. Phys.: Condens. Matter}\ }\textbf
  {\bibinfo {volume} {17}},\ \bibinfo {pages} {S3481} (\bibinfo {year}
  {2005})}\BibitemShut {NoStop}%
\bibitem [{\citenamefont {Marconi}\ and\ \citenamefont
  {Tarazona}(1999)}]{Marconi1999}%
  \BibitemOpen
  \bibfield  {author} {\bibinfo {author} {\bibfnamefont {U.~M.~B.}\
  \bibnamefont {Marconi}}\ and\ \bibinfo {author} {\bibfnamefont
  {P.}~\bibnamefont {Tarazona}},\ }\href {\doibase 10.1063/1.478705} {\bibfield
   {journal} {\bibinfo  {journal} {J. Chem. Phys.}\ }\textbf {\bibinfo {volume}
  {110}},\ \bibinfo {pages} {8032} (\bibinfo {year} {1999})}\BibitemShut
  {NoStop}%
\bibitem [{\citenamefont {Schmidt}\ and\ \citenamefont
  {Brader}(2013)}]{Schmidt2013}%
  \BibitemOpen
  \bibfield  {author} {\bibinfo {author} {\bibfnamefont {M.}~\bibnamefont
  {Schmidt}}\ and\ \bibinfo {author} {\bibfnamefont {J.~M.}\ \bibnamefont
  {Brader}},\ }\href {\doibase 10.1063/1.4807586} {\bibfield  {journal}
  {\bibinfo  {journal} {J. Chem. Phys.}\ }\textbf {\bibinfo {volume} {138}},\
  \bibinfo {pages} {214101} (\bibinfo {year} {2013})}\BibitemShut {NoStop}%
\bibitem [{\citenamefont {te~Vrugt}, \citenamefont {Löwen},\ and\
  \citenamefont {Wittkowski}(2020)}]{teVrugt2020}%
  \BibitemOpen
  \bibfield  {author} {\bibinfo {author} {\bibfnamefont {M.}~\bibnamefont
  {te~Vrugt}}, \bibinfo {author} {\bibfnamefont {H.}~\bibnamefont {Löwen}}, \
  and\ \bibinfo {author} {\bibfnamefont {R.}~\bibnamefont {Wittkowski}},\
  }\href {\doibase 10.1080/00018732.2020.1854965} {\bibfield  {journal}
  {\bibinfo  {journal} {Adv. Phys.}\ }\textbf {\bibinfo {volume} {69}},\
  \bibinfo {pages} {121} (\bibinfo {year} {2020})}\BibitemShut {NoStop}%
\bibitem [{\citenamefont {\relax Wolfram Research{,}~Inc.}()}]{mathematica}%
  \BibitemOpen
  \bibfield  {author} {\bibinfo {author} {\bibnamefont {\relax Wolfram
  Research{,}~Inc.}},\ }\href {https://www.wolfram.com/mathematica} {\enquote
  {\bibinfo {title} {Mathematica, {V}ersion 11.3},}\ }\bibinfo {note}
  {Champaign, IL, 2020}\BibitemShut {NoStop}%
\bibitem [{\citenamefont {Johnson}()}]{Johnson2017}%
  \BibitemOpen
  \bibfield  {author} {\bibinfo {author} {\bibfnamefont {S.~G.}\ \bibnamefont
  {Johnson}},\ }\href@noop {} {}\bibinfo {note} {Cubature Package,
  \url{https://github.com/stevengj/cubature} (accessed Oct. 15,
  2022)}\BibitemShut {NoStop}%
\end{thebibliography}%

\end{document}